\documentclass[11pt,a4paper]{article} 
\usepackage{jheppub}
\usepackage[utf8]{inputenc}

\usepackage{graphicx}
\usepackage{amsmath}
\usepackage{amsfonts}
\newcommand{\beq}{\begin{equation}}
\newcommand{\eeq}{\end{equation}}
\newcommand{\beqa}{\begin{eqnarray}}
\newcommand{\eeqa}{\end{eqnarray}}
\def\lsim{\raise0.3ex\hbox{$<$\kern-0.75em\raise-1.1ex\hbox{$\sim$}}}
\def\gsim{\raise0.3ex\hbox{$>$\kern-0.75em\raise-1.1ex\hbox{$\sim$}}}

\def\0{{\bf 0}}

\def\kk{{\kappa}}
\def\pp{{\hat{p}}}

\title{Event-shape engineering and heavy-flavour observables in relativistic heavy-ion collisions}

\author[a]{Andrea Beraudo}
\author[a]{Arturo De Pace}
\author[a]{Marco Monteno}
\author[a]{Marzia Nardi}
\author[a]{Francesco Prino}

\affiliation[a]{INFN, Sezione di Torino, via Pietro Giuria 1, I-10125 Torino}

\emailAdd{beraudo@to.infn.it}
\emailAdd{depace@to.infn.it}
\emailAdd{monteno@to.infn.it}
\emailAdd{nardi@to.infn.it}
\emailAdd{prino@to.infn.it}

\abstract{Traditionally, events collected at relativistic heavy-ion colliders are classified according to some centrality estimator (e.g. the number of produced charged particles) related to the initial energy density and volume of the system. In a naive picture the latter are directly related to the impact parameter of the two nuclei, which sets also the initial eccentricity of the system: zero in the case of the most central events and getting larger for more peripheral collisions. A more realistic modelling requires to take into account event-by-event fluctuations, in particular in the nucleon positions within the colliding nuclei: collisions belonging to the same centrality class can give rise to systems with different initial eccentricity and hence different flow harmonics for the final hadron distributions. This issue can be addressed by an event-shape-engineering analysis, consisting in selecting events with the same centrality but different magnitude of the average bulk anisotropic flow and therefore of the initial-state eccentricity. In this paper we present the implementation of this analysis in the POWLANG transport model, providing predictions for the transverse-momentum and angular distributions of charm and beauty hadrons for event-shape selected collisions. In this way it is possible to get information on how the heavy quarks propagating (and hadronizing) in a hot environment respond both to its energy density and to its geometric asymmetry, breaking the perfect correlation between eccentricity and impact parameter which characterizes a modelling of the medium based on smooth average initial conditions.}

\begin{document}

\maketitle

\section{Introduction}
Heavy flavour particles ($D/B$ mesons and $\Lambda_{c/b}$ baryons), arising from charm and beauty quarks produced in initial hard partonic scattering processes, have been always considered a probe of the deconfined medium one expects to form in relativistic heavy-ion collisions. The scope of first heavy-flavour measurements was simply to understand whether, in spite of the large mass of the parent quarks, the distributions of final-state particles displayed the same features observed in the case of light hadrons, i.e. a quenching of the spectra at high transverse momentum $p_T$ (with possible signatures of a mass and colour-charge dependence of parton energy loss) and a non-vanishing, positive elliptic-flow coefficient $v_2$. First data were limited to electrons from heavy-flavour decays, without the possibility to discriminate between the charm and beauty contributions~\cite{Adare:2010de,Abelev:2006db}. It became then possible to reconstruct $D$-mesons through some exclusive decay channels~\cite{ALICE:2012ab,Adamczyk:2014uip,Sirunyan:2017xss,Abelev:2013lca,Adamczyk:2017xur,Sirunyan:2017plt}. The message from these first measurements was that, although quantitatively a bit milder, the same quenching of the momentum spectra and elliptic (and triangular, as shown in Ref.~\cite{Sirunyan:2017plt}) asymmetry of the azimuthal distributions observed for light hadrons characterized also charm and beauty particles. This entailed a quite strong coupling of the heavy quarks with the hot deconfined plasma of quarks and gluons (QGP) supposed to be produced in the collision of the two nuclei and possibly a non trivial modification of their hadronization due to the large density of light thermal partons nearby. From a comparison of the outcomes of transport calculations with experimental data it is in principle possible to extract information on the value of the heavy-quark momentum-diffusion coefficient, a fundamental quantity which in the static limit in hot-QCD admits a rigorous definition in terms of Euclidean correlators of chromo-electric fields~\cite{Francis:2015daa,CaronHuot:2008uh}. Recently, a systematic investigation based on a Bayesian approach aiming at extracting the heavy-flavour diffusion coefficient from current experimental data has been carried out by some authors~\cite{Xu:2017hgt}. In this connection a comprehensive study of the various theoretical uncertainties arising from the initial heavy-quark spectrum, from Cold-Nuclear-Matter effects and from the modelling of the medium and of hadronization was carried out in Ref.~\cite{Rapp:2018qla}. This is somehow similar to what done for the case of soft hadrons, where a comparison of hydrodynamics calculations with experimental particle distributions allowed people to constrain within a quite narrow band another transport coefficient, the shear-viscosity to entropy-density ratio $\eta/s$, which turned out to be close to the lower bound $1/4\pi$ postulated by the AdS/CFT correspondence~\cite{Policastro:2001yc}.

More recent measurements opened the possibility to get access to a richer information. Studies of $D_s$ and $\Lambda_c$ production in nuclear collisions have the potential to put the issue of medium-modification of heavy-flavour hadrochemistry on solid ground~\cite{Adam:2015jda,Zhou:2017ikn}, making possible to validate heavy-quark hadronization models based on the recombination with light thermal partons. Experimental data on $B$ meson production~\cite{Sirunyan:2017oug} allow one to study the mass dependence of the heavy-quark medium interaction; if in the future these analysis were extended to lower transverse momentum they would allow one to perform a theory-to-experiment comparison in a kinematic region in which transport calculations are under the best control, reaching the goal of really \emph{measuring} the heavy-quark diffusion coefficient. Recently heavy-flavour studies have been extended to the case of proton-nucleus collisions~\cite{Adare:2012yxa,Abelev:2014hha,Adam:2016ich,Sirunyan:2018toe}, with the aim of contributing to answer the still open question whether also in such small systems QGP droplets can be formed~\cite{Beraudo:2015wsd}.

Finally, the measurement of odd flow-harmonics of heavy-flavour hadrons can provide a richer information on the initial conditions of the system formed after the collision of the two high-energy nuclei, like its tilted profile in the reaction plane (wounded nucleons tending to deposit more energy along the direction of their motion) in the case of the directed flow $v_1$~\cite{Singha:2018cdj,Chatterjee:2017ahy,Das:2016cwd} or its event-by-event fluctuations (from the random nucleon positions) in the case of the triangular flow $v_3$. The triangular flow $v_3$ of $D$ mesons in Pb-Pb collisions provided by transport calculations has been studied in some recent publications and theoretical results~\cite{Nahrgang:2014vza,Beraudo:2017gxw} have been compared to experimental data from the CMS collaboration~\cite{Sirunyan:2017plt}.

A further possibility of accessing the response of the final particle distributions to the initial asymmetries of the system, getting information both on the coupling of the heavy quarks with the medium and on its initial conditions, is given by the so-called Event-Shape-Engineering (ESE) studies. The basic idea is to select events belonging to the same centrality class, but characterized by a different initial geometric (elliptic or triangular) asymmetry, getting subsamples of events with high/low eccentricity~\cite{Schukraft:2012ah}. Such an approach was proposed and adopted by the ALICE collaboration in the analysis of momentum and azimuthal distributions of light hadrons~\cite{Adam:2015eta}, comparing the results obtained in subsamples of collisions with large/small average elliptic-flow with the ones of an unbiased selection of events. Here, in the framework of a transport calculation, we wish to extend the approach to heavy flavour, studying how the different geometric asymmetry and the resulting anisotropic flow of the medium affect the propagation of heavy quarks and leave their signatures in the final charm/beauty-hadron distributions~\cite{Beraudo:2018bxb}. Our findings will be compared with recent experimental outcomes~\cite{Acharya:2018bxo}. For independent phenomenological studies of hard probes (heavy-flavour particles and jets) in heavy-ion collisions based on event-shape-engineering see also Refs.~\cite{Prado:2016szr,Christiansen:2016uaq,Gossiaux:2017nwz}.

Our paper is organized as follows. In Sec.~\ref{Sec:medium} we present our modelling of the background medium, focusing on the simulation of the initial conditions, on the selection -- in the various centrality classes -- of the events belonging to different eccentricity subsamples and on the resulting light-hadron spectra decoupling from the fireball at the end of its hydrodynamic evolution. In Sec.~\ref{Sec:transport} we briefly summarize our setup for the simulation of heavy-quark transport and hadronization. In Sec.~\ref{Sec:results} we display the results of our transport calculations performed with the POWLANG model, which account both for the propagation of $c$ and $b$ quarks through the QGP and for their hadronization in the presence of a hot deconfined medium. Finally in Sec.~\ref{Sec:conclusions} we discuss our results, suggesting possible future improvements.

\section{Modelling of the background medium}\label{Sec:medium}
For the modelling of the medium produced in nucleus-nucleus collisions (in this paper we consider Pb-Pb collisions at $\sqrt{s_{\rm NN}}\!=\!5.02$ TeV) we adopted the same approach described in detail in Ref.~\cite{Beraudo:2017gxw}, interfacing a Glauber Monte-Carlo (Glauber-MC) simulation of the initial condition of the system to a hydrodynamic code (ECHO-QGP~\cite{DelZanna:2013eua}) calculating the subsequent evolution of the matter, under the assumption of longitudinal boost-invariance; the latter is a good approximation for observables around mid-rapidity and allows one to solve a (2+1)-dimensional problem, reducing the computational time.

In order to set the initial geometry we distribute nucleons within the two nuclei according to a Woods-Saxon distribution and we generate several thousands ($\sim 30000$, to have a sufficient statistics) of Pb-Pb collisions at random impact parameter organizing them in centrality classes according to the number of binary nucleon-nucleon collisions: a nucleon-nucleon inelastic cross-section $\sigma_{\rm NN}^{\rm in}\!=\!70$ mb was employed in the simulation. For a given event each nucleon-nucleon collision is taken as a source of entropy production, with a Gaussian smearing $\sigma$. The initial entropy density in the transverse plane used to start the hydrodynamic evolution of the system at the longitudinal proper time $\tau_0\!=\!0.5$ fm/c reads then
\beq
s(x,y)=\frac{K}{2\pi\sigma^2}\sum_{i=1}^{N_{\rm coll}}\exp\left[-\frac{(x-x_i)^2+(y-y_i)^2}{2\sigma^2}\right].\label{eq:incond}
\eeq
The parameter $K$ (with dimensions of an inverse length) sets the average entropy deposited by a single collision (so far we do not include fluctuations at the level of the individual nucleon-nucleon inelastic collisions). As in Ref.~\cite{Beraudo:2017gxw} for Pb-Pb collisions at $\sqrt{s_{\rm NN}}\!=\!5.02$ TeV we choose $K\tau_0\!=\!6.37$. 
For each event the above entropy density can be used as a weight to define complex eccentricities, which characterize the initial state (i.e. both the amount of anisotropy and its orientation in the transverse plane) and will be mapped into the final hadron distributions by the subsequent hydrodynamic evolution~\cite{Qiu:2011iv}:
\beq
{\epsilon_m}e^{im{\Psi_m}}\equiv-\frac{\left\{r_\perp^2e^{im\phi}\right\}}{\{r_\perp^2\}},\quad{\rm with}\quad\{...\}\equiv\int \!d^2r_\perp\, s(\vec r_\perp)(...).
\eeq
Modulus and orientation of the various azimuthal harmonics are given by:
\beqa
\epsilon_{m}&=&\frac{\sqrt{\{r_\perp^2\cos(m\phi)\}^2+\{r_\perp^2\sin(m\phi)\}^2}}{\{r_\perp^2\}}\\
\Psi_{m}&=&\frac{1}{m}\,{\rm atan2}\left(-\{r_\perp^2\sin(m\phi)\},-\{r_\perp^2\cos(m\phi)\}\right)
\eeqa

\begin{figure}[!ht]
\begin{center}
\includegraphics[clip,width=0.48\textwidth]{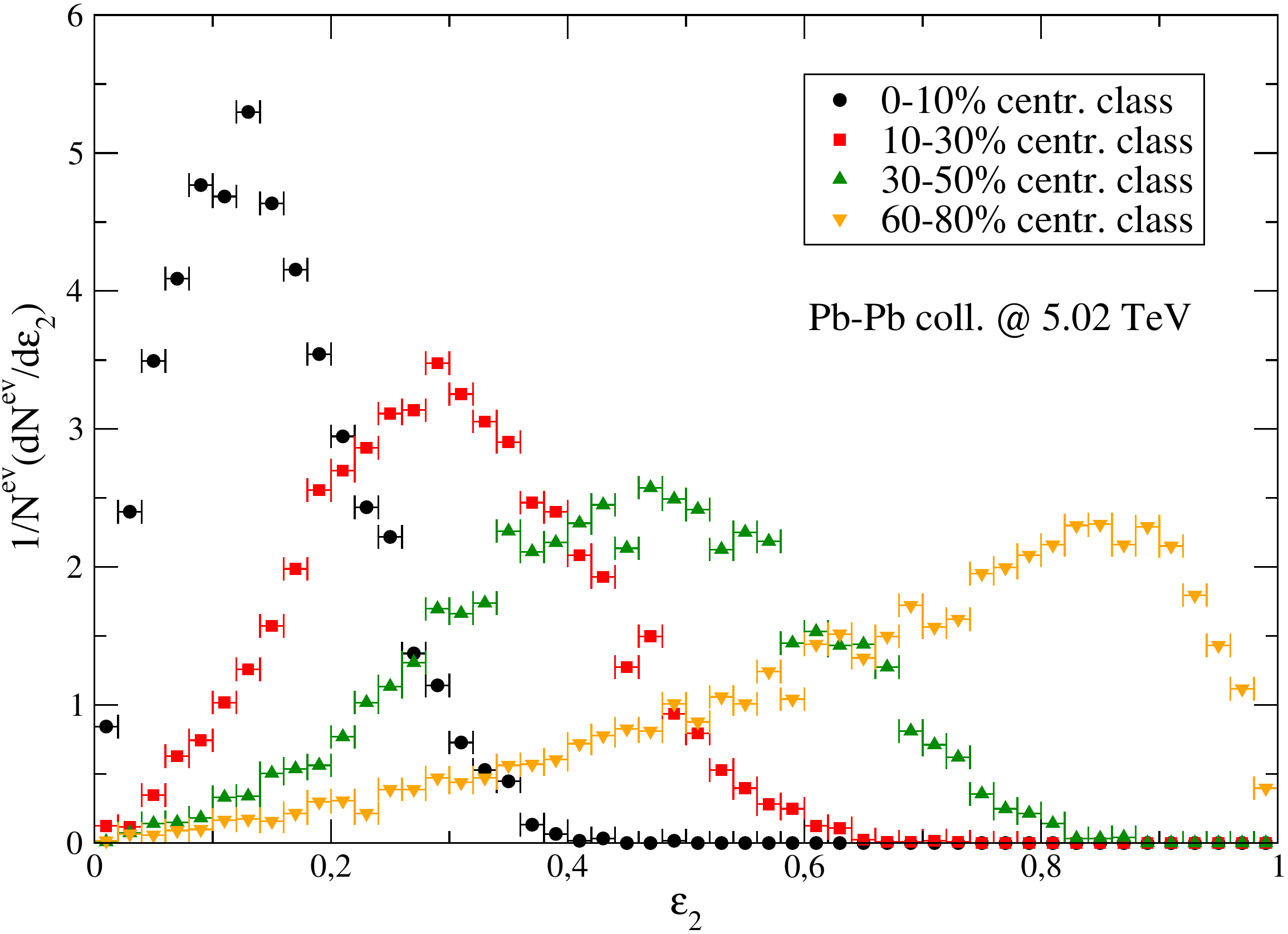}
\includegraphics[clip,width=0.48\textwidth]{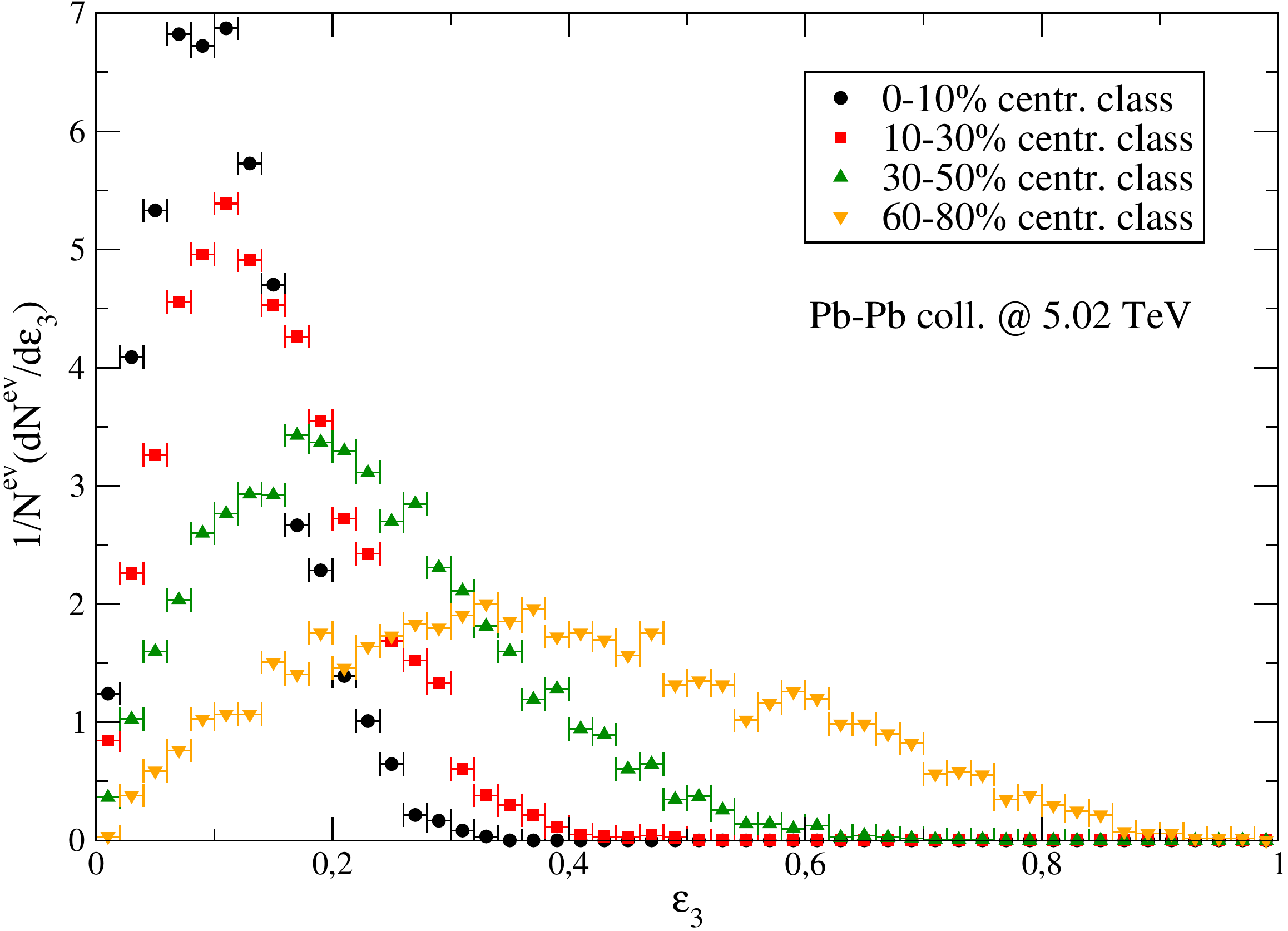}
\caption{The elliptic (left panel) and triangular (right panel) eccentricity distribution of Pb-Pb collisions at $\sqrt{s_{\rm NN}}\!=\!5.02$ TeV belonging to different centrality classes.}\label{fig:dNdeps} 
\end{center}
\end{figure}
\begin{figure}[!ht]
\begin{center}
\includegraphics[clip,width=0.48\textwidth]{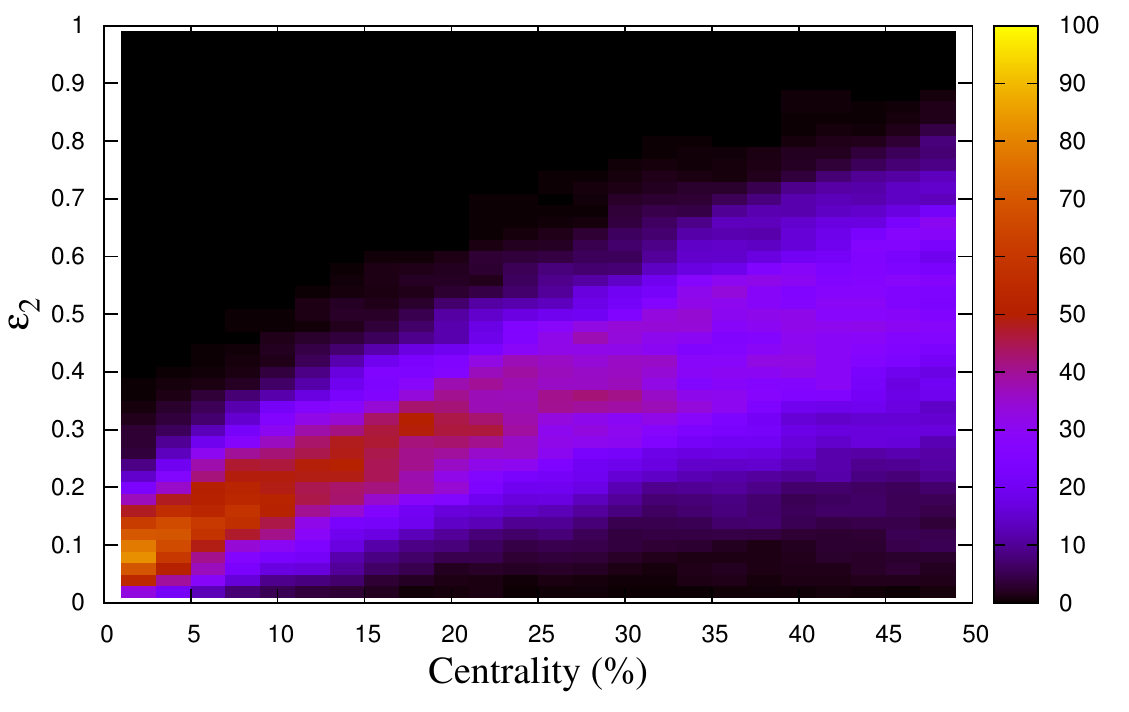}
\includegraphics[clip,width=0.48\textwidth]{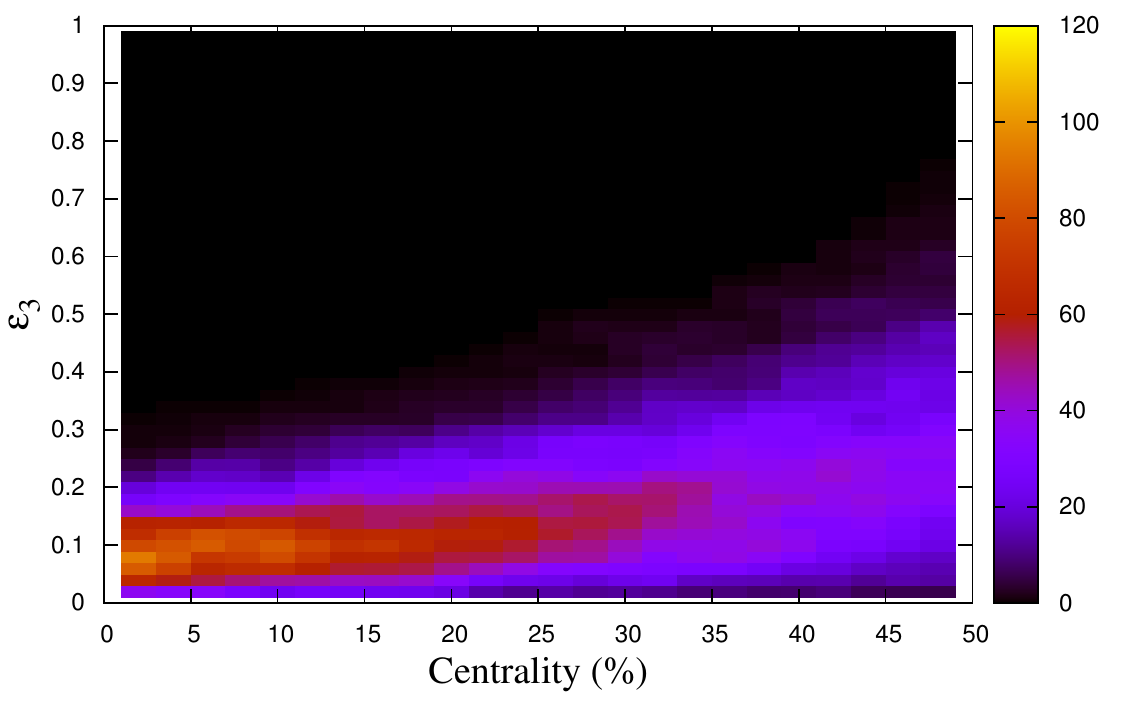}
\caption{Correlation between the elliptic (left panel) and triangular (right panel) eccentricity and the centrality of the nucleus-nucleus collisions for our Glauber-MC sample of Pb-Pb events.}\label{fig:ecc-centr_corr}
\end{center}
\end{figure}
Using as an estimator the number of binary nucleon-nucleon collisions, we group the Pb-Pb events in centrality classes (0-10\%, 10-30\% and 30-50\%) and study within each sample the distribution of initial elliptic and triangular eccentricity $\epsilon_2$ and $\epsilon_3$. We will also consider in some of the  calculations a very peripheral class (60-80\%). Results are shown in Fig.~\ref{fig:dNdeps}. Notice how, within a given centrality class, the eccentricity distribution is quite broad, in particular for the case of $\epsilon_2$ whose large event-by-event fluctuations arise both from the different impact parameter and from the random positions of the nucleons within the colliding nuclei. The strong dependence on the impact parameter is also evident from the sizable shift of the peak of the distribution towards larger values of $\epsilon_2$ going from central to peripheral collisions. On the other hand, in the case of $\epsilon_3$ the eccentricity distributions are narrower and the displacement of the peak when moving to a different centrality class is milder. This reflects the different origin of the triangular asymmetry, which (neglecting sub-nucleonic degrees of freedom) is entirely due to the event-by-event fluctuations in the positions of the nucleons inside the colliding nuclei.

In order to study how the initial asymmetry of the system is mapped by the hydrodynamic/transport evolution into the azimuthal anisotropies of the final particle distributions (both light and heavy-flavour hadrons, the latter being the focus of this work) we select, for each centrality class, the 20\% most eccentric and the 60\% least eccentric events. This corresponds to the selections adopted by the ALICE collaboration in the recent heavy-flavour analysis in Ref.~\cite{Acharya:2018bxo}. We do this both for $\epsilon_2$ and $\epsilon_3$. This, depending on the cases, amounts to subsamples of several hundreds/thousands of events. As evident from Fig.~\ref{fig:ecc-centr_corr} there is a strong anti-correlation between the elliptic eccentricity $\epsilon_2$ and the centrality of the collision (the effect is much milder for the case of $\epsilon_3$).
In order to quantify the effect it is useful to provide some typical numbers. For the 0-10\% centrality class one has, for an unbiased selection of events, $\langle N_{\rm coll}\rangle_{N_{\rm coll}}^{\rm unbias.}\!=\!1658$ (in the average, events are weighted by $N_{\rm coll}$, since heavy-quark production scales with the number of binary collisions); applying cuts on elliptic and triangular eccentricity one gets $\langle N_{\rm coll}\rangle_{N_{\rm coll}}^{{\rm high}-\epsilon_2}\!=\!1471$, $\langle N_{\rm coll}\rangle_{N_{\rm coll}}^{{\rm low}-\epsilon_2}\!=\!1730$ and $\langle N_{\rm coll}\rangle_{N_{\rm coll}}^{{\rm high}-\epsilon_3}\!=\!1604$, $\langle N_{\rm coll}\rangle_{N_{\rm coll}}^{{\rm low}-\epsilon_3}\!=\!1675$, respectively.
Hence, selecting events within a given centrality class of higher/lower eccentricity leads to a sample biased toward lower/higher centrality. We have to bear in mind this observation in interpreting our numerical findings. Experimental analysis, profiting from a huge statistics, remove this bias performing their selection on eccentricity in very small bins of centrality.

\begin{figure}[!ht]
\begin{center}
\includegraphics[clip,width=0.32\textwidth]{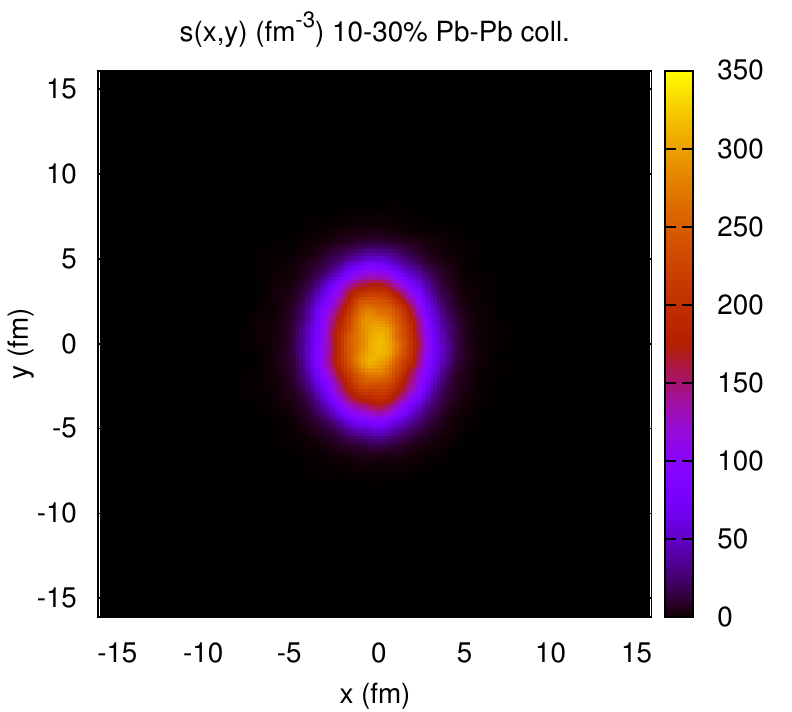}
\includegraphics[clip,width=0.32\textwidth]{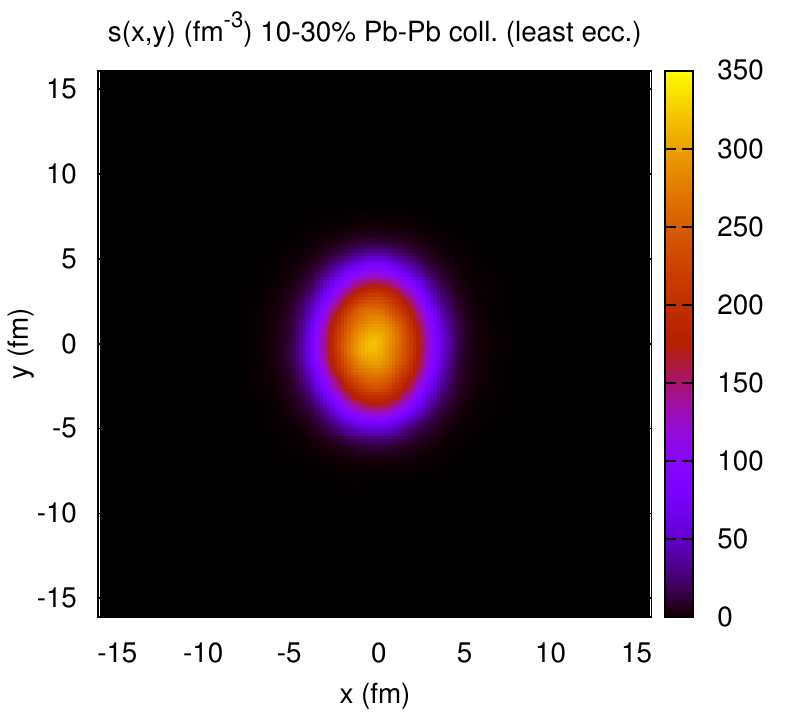}
\includegraphics[clip,width=0.32\textwidth]{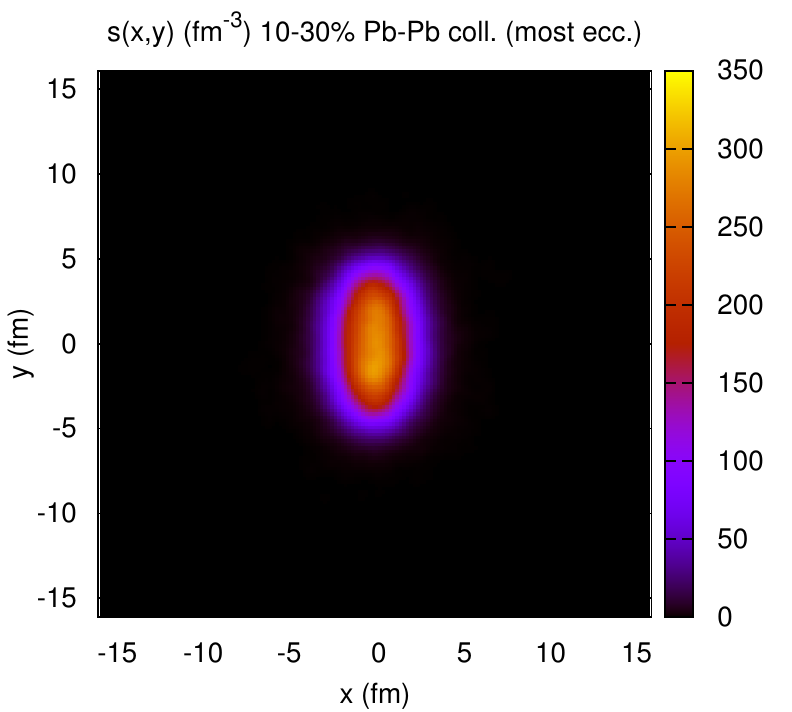}
\caption{The initial entropy-density for the 10-30\% most central Pb-Pb collisions at $\sqrt{s_{\rm NN}}\!=\!5.02$ TeV. The three panels refer to no eccentricity cut (left), the 60\% of events with the lowest elliptic deformation $\epsilon_2$ (middle) and the 20\% of events with the highest $\epsilon_2$ (right).}\label{fig:incond-e2_10-30} 
\end{center}
\end{figure}
\begin{figure}[!ht]
\begin{center}
\includegraphics[clip,width=0.32\textwidth]{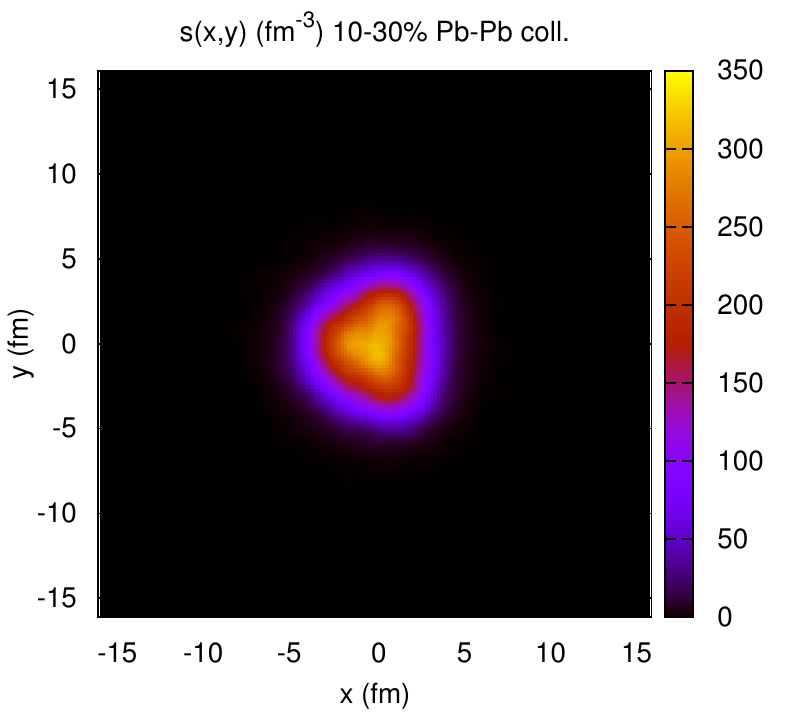}
\includegraphics[clip,width=0.32\textwidth]{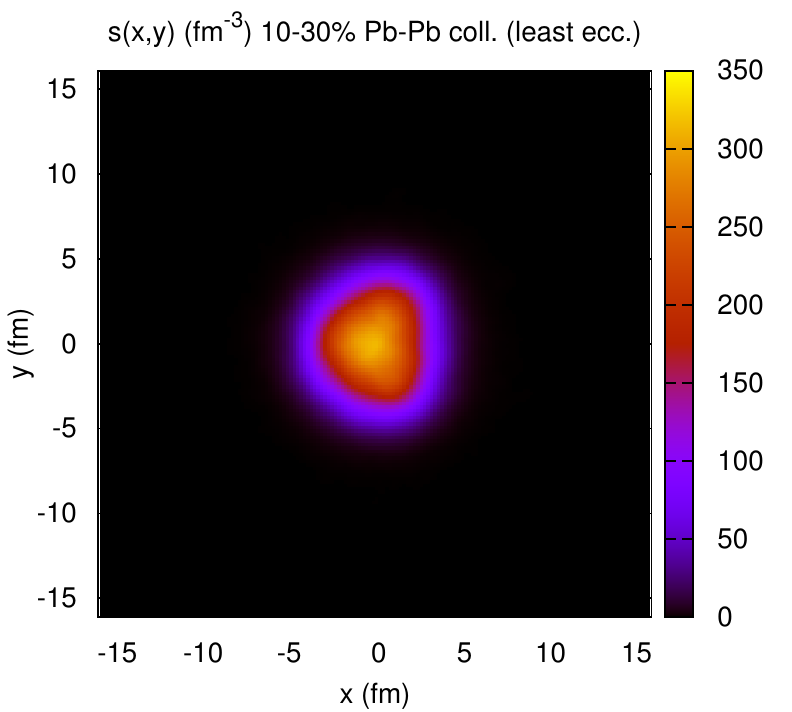}
\includegraphics[clip,width=0.32\textwidth]{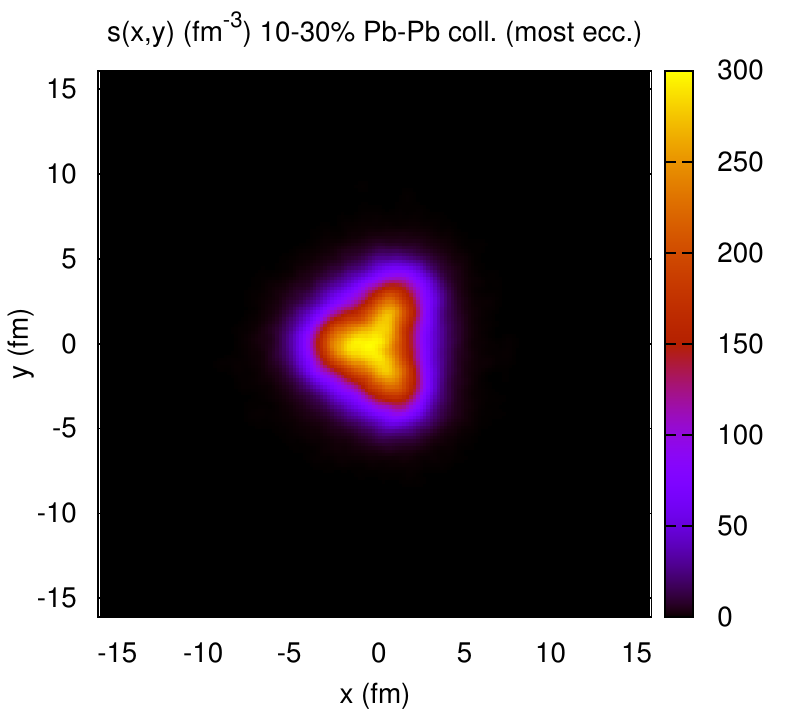}
\caption{The initial entropy-density for the 10-30\% most central Pb-Pb collisions at $\sqrt{s_{\rm NN}}\!=\!5.02$ TeV. The three panels refer to no eccentricity cut (left), the 60\% of events with the lowest triangular deformation $\epsilon_3$ (middle) and the 20\% of events with the highest $\epsilon_3$ (right).}\label{fig:incond-e3_10-30} 
\end{center}
\end{figure}
The Glauber-MC modelling of the initial state can now be used as the initial condition of the hydrodynamic evolution of the system, which we describe through the ECHO-QGP code~\cite{DelZanna:2013eua}. Its output provides the information on the background medium through which the propagation of the heavy quarks takes place.
If we wished to perform fully realistic simulations we should numerically solve the set of hydrodynamic equations for all the ($\sim 15000$ if we focus on the 50\% most central events) different initial condition, simulating then the Langevin evolution of the heavy quarks for each of these events. This would require huge computing and storage resources. Therefore, we decided to follow the approach adopted in Refs.~\cite{Beraudo:2015wsd,Beraudo:2017gxw}, producing for each of the subsamples of events of interest an average initial condition which embeds the effect of the fluctuations one wants to analyze.
Within a given centrality class we collect the subsample of events satisfying the cut on eccentricity (i.e. low-$\epsilon_n$, high-$\epsilon_n$, unbiased if no cut is applied), we rotate each of them so that its relevant symmetry plane $\psi_n$ is aligned along the $x$-axis and, starting from Eq.~(\ref{eq:incond}), we construct an average entropy-density distribution weighting each event by the number of binary nucleon-nucleon collisions (since the $Q\overline Q$ production scales with $N_{\rm coll}$, which introduces a bias towards more central events). 
In Figs.~\ref{fig:incond-e2_10-30} and~\ref{fig:incond-e3_10-30}, referring to the 10-30\% centrality class,  we display the result of such a procedure for the study of the response to an elliptic and triangular deformation, respectively: the average initial conditions for the unbiased, low-$\epsilon_n$ and high-$\epsilon_n$ subsets of events are shown. 

\begin{figure}[!ht]
\begin{center}
\includegraphics[clip,width=0.48\textwidth]{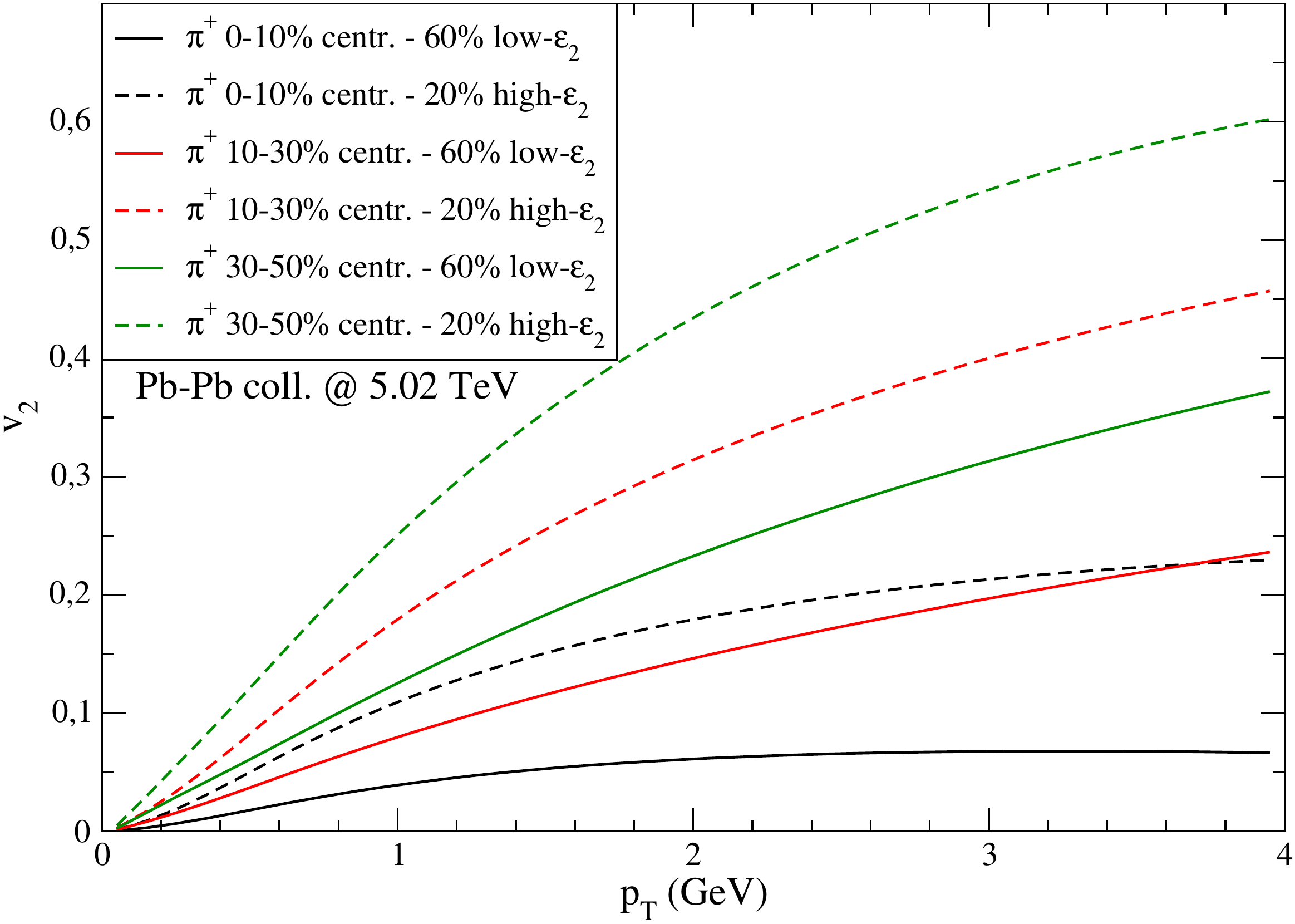}
\includegraphics[clip,width=0.48\textwidth]{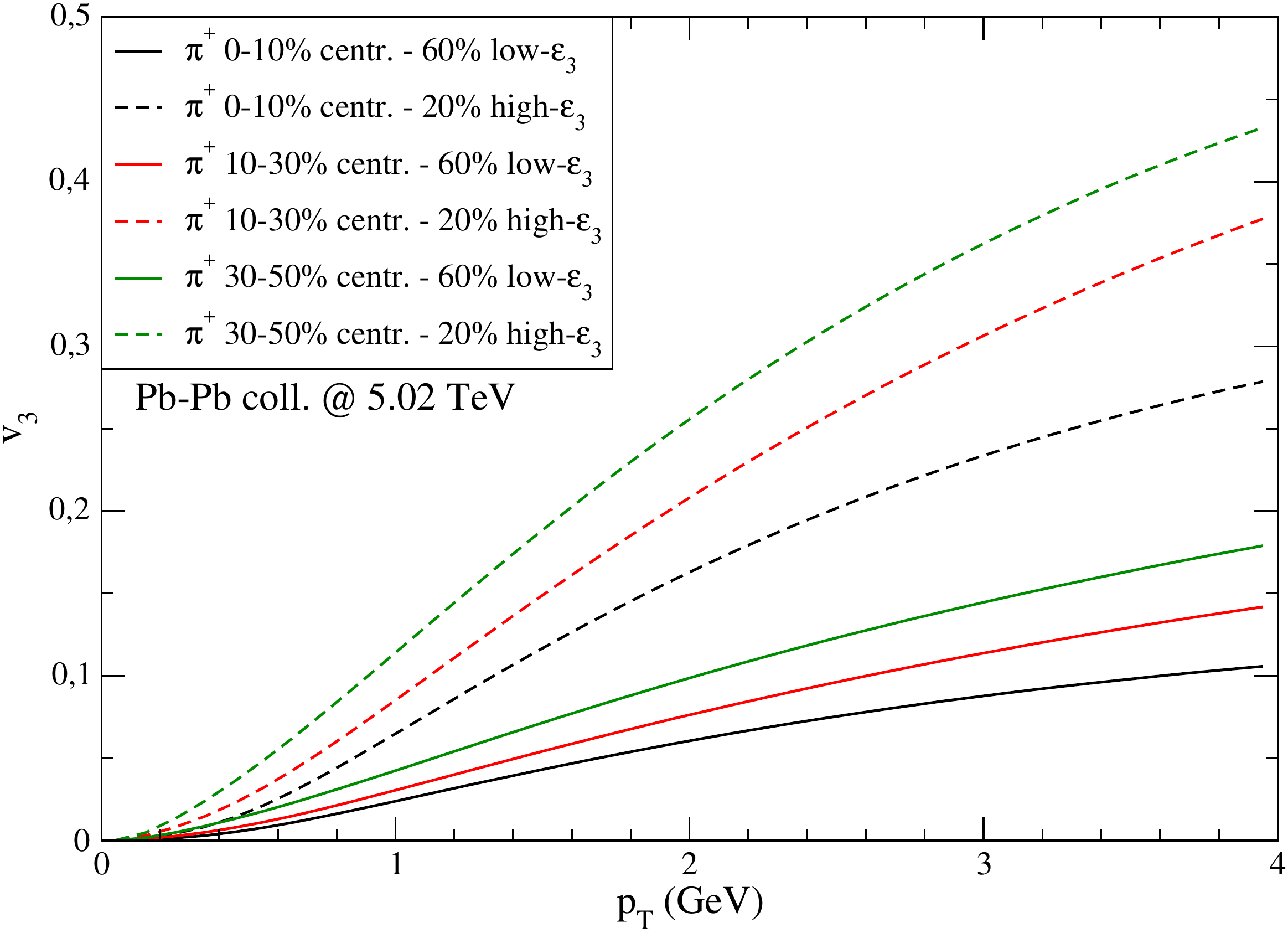}
\caption{The elliptic (left panel) and triangular (right panel) flow of charged pions in Pb-Pb collisions at $\sqrt{s_{\rm NN}}\!=\!5.02$ TeV for different eccentricity selections within the same centrality class.}\label{fig:vn_light-validation} 
\end{center}
\end{figure}
\begin{figure}[!ht]
\begin{center}
  \includegraphics[clip,width=0.48\textwidth]{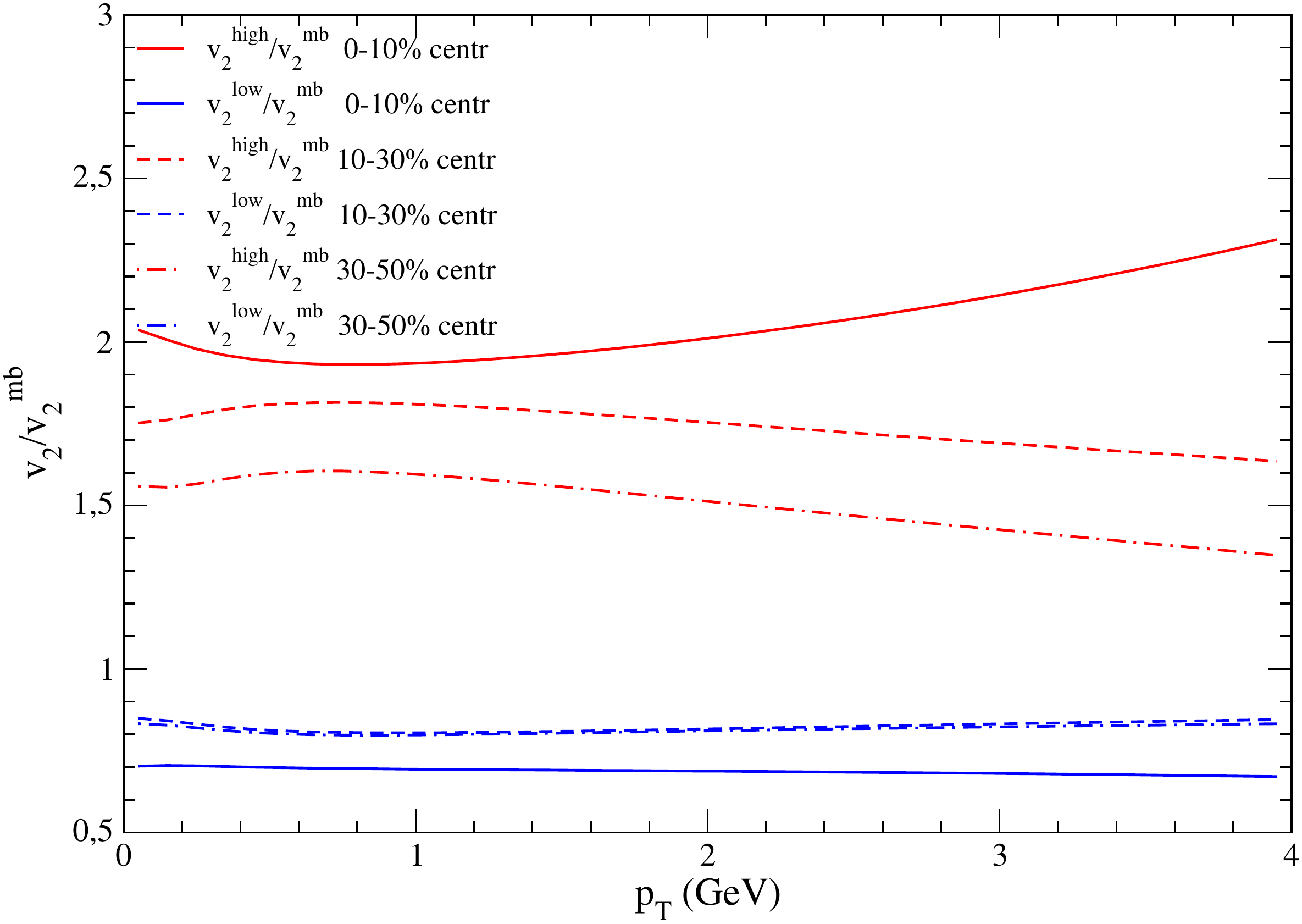}
  \includegraphics[clip,width=0.48\textwidth]{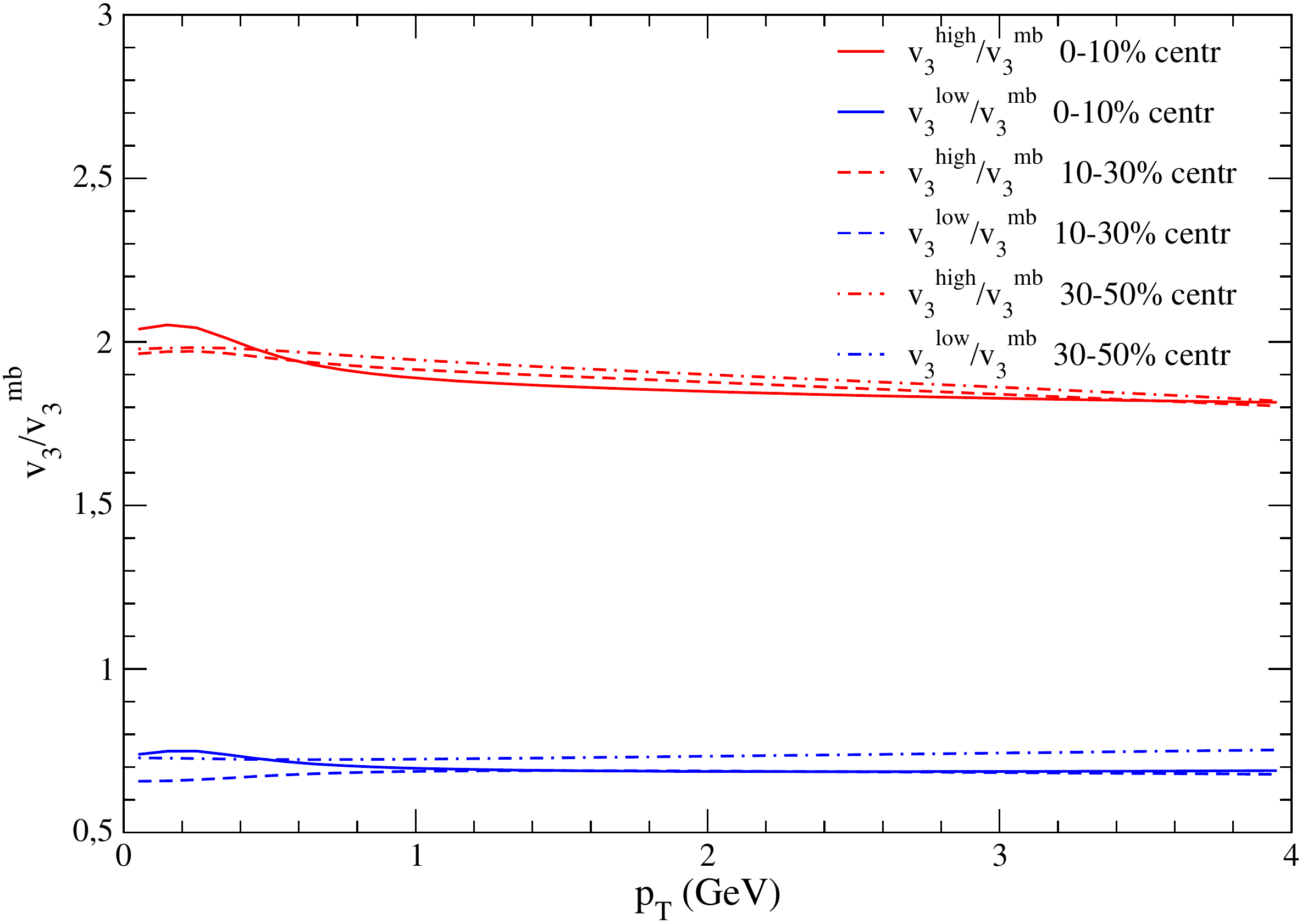}  
\caption{Ratio of the charged-pion elliptic (left) and triangular (right) flow of high/low-eccentricity events over the minimum-bias ones within the same centrality class. Results of our hydrodynamic modelling can be compared to ALICE data from Ref.~\cite{Adam:2015eta}.}\label{fig:v2vsv2mb} 
\end{center}
\end{figure}
Having performed the hydrodynamic evolution of the medium we can check the effect of the eccentricity fluctuations on the light-hadron distributions, obtained from a standard Cooper-Frye decoupling from a freeze-out hypersurface. In Fig.~\ref{fig:vn_light-validation} we display the resulting elliptic and triangular flow coefficients $v_2$ and $v_3$ for charged pions, defined as $v_n=\langle\cos[n(\phi-\Psi_n)]\rangle$. Notice how selecting events with high/low eccentricity produces a huge effect on the final angular distributions, comparable or even larger than changing centrality class, this in particular for the case of $v_3$. It is of interest to quantify the effect by taking the ratio of the elliptic flow in the high/low-$\epsilon_2$ subsets over the one in the unbiased sample. Our results are displayed in Fig.~\ref{fig:v2vsv2mb}. The effect of the event-shape selection on the $v_2$ coefficient appears in qualitative agreement with recent ALICE data~\cite{Adam:2015eta}, although obtained with different eccentricity cuts. Notice that our results for the $v_2$ in eccentricity-selected events display some dependence on centrality. This probably reflects our procedure of selection, which does not decouple completely eccentricity from centrality. On the other hand in the case of $v_3$, in which the initial geometric deformation arises entirely from fluctuations in the nucleon positions uncorrelated with the impact parameter, the results in the right panel of Fig.~\ref{fig:v2vsv2mb} do not show any dependence on centrality.

\begin{figure}[!ht]
\begin{center}
\includegraphics[clip,width=0.32\textwidth]{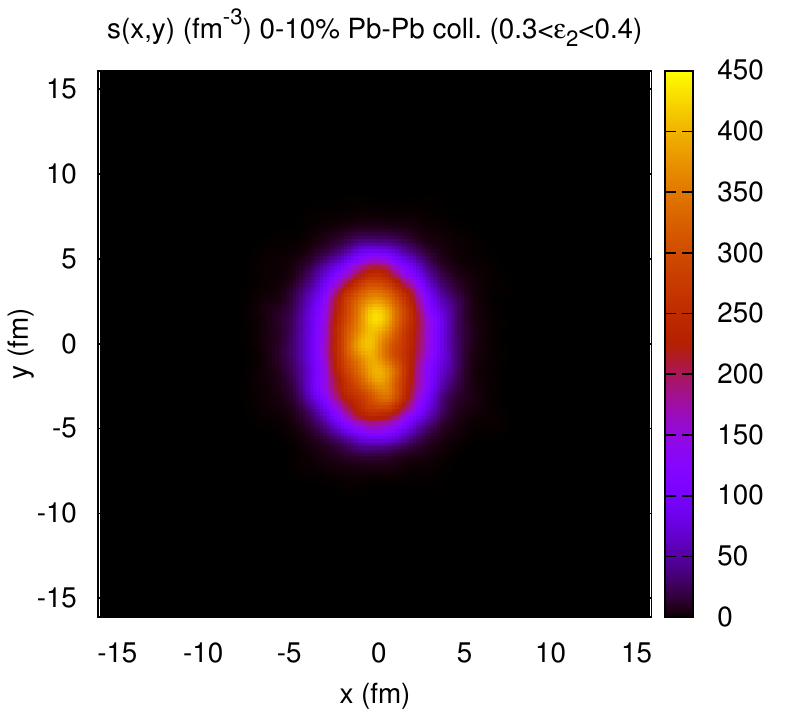}
\includegraphics[clip,width=0.32\textwidth]{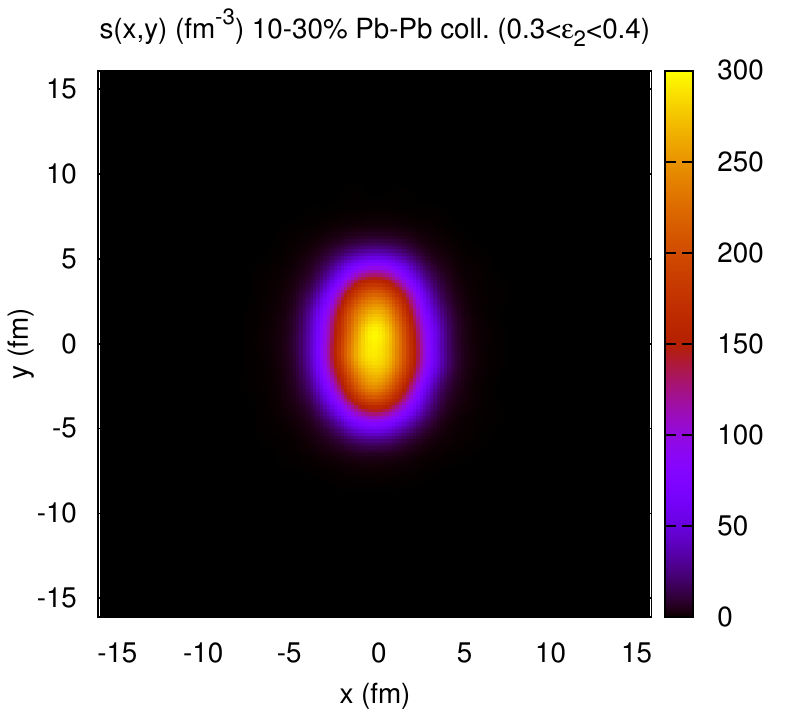}
\includegraphics[clip,width=0.32\textwidth]{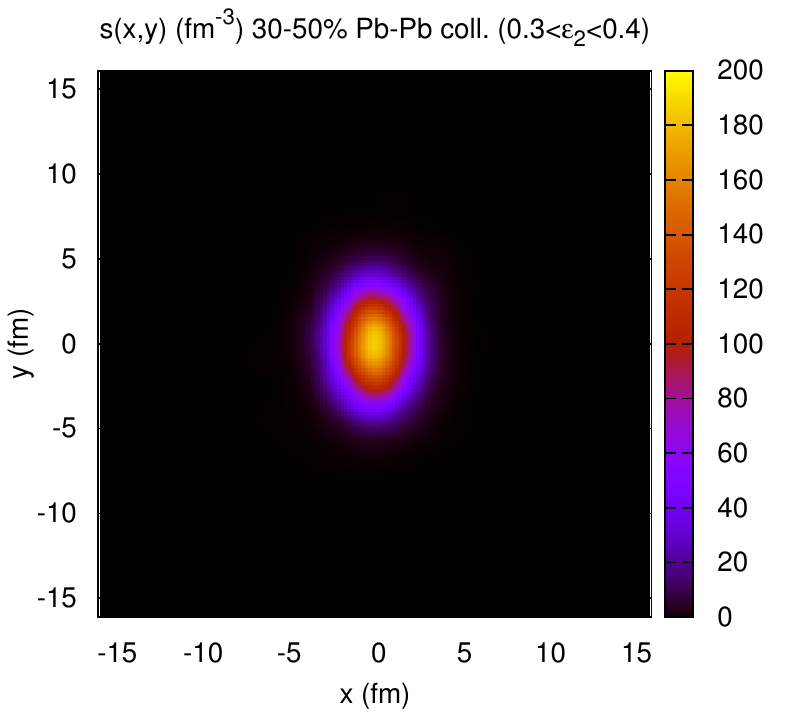}
\caption{The initial entropy-density for Pb-Pb collisions at $\sqrt{s_{\rm NN}}\!=\!5.02$ TeV of fixed elliptic eccentricity $0.3\le\epsilon_2\le 0.4$ and different centrality classes.}\label{fig:incond-e2fixed} 
\end{center}
\end{figure}
\begin{figure}[!ht]
\begin{center}
\includegraphics[clip,width=0.32\textwidth]{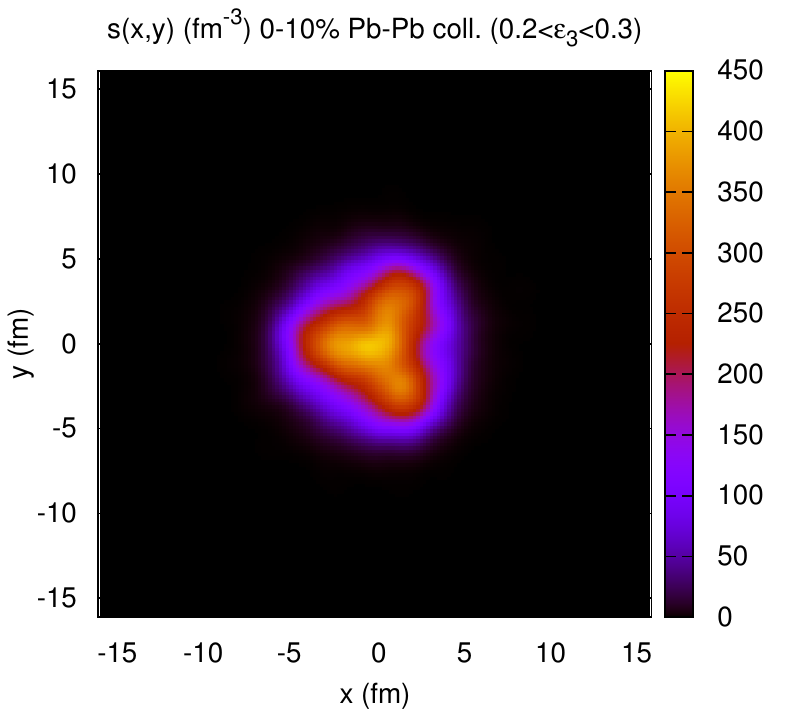}
\includegraphics[clip,width=0.32\textwidth]{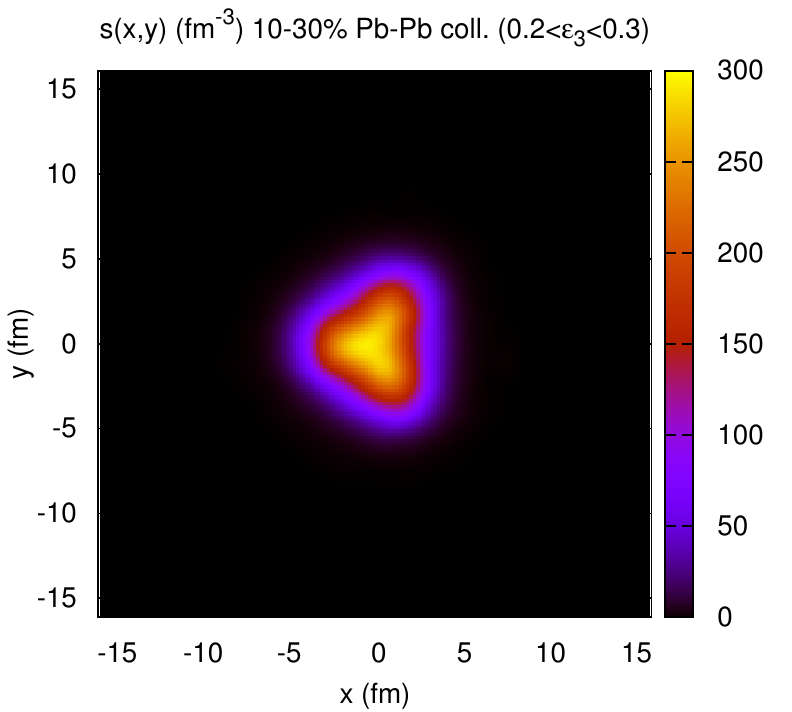}
\includegraphics[clip,width=0.32\textwidth]{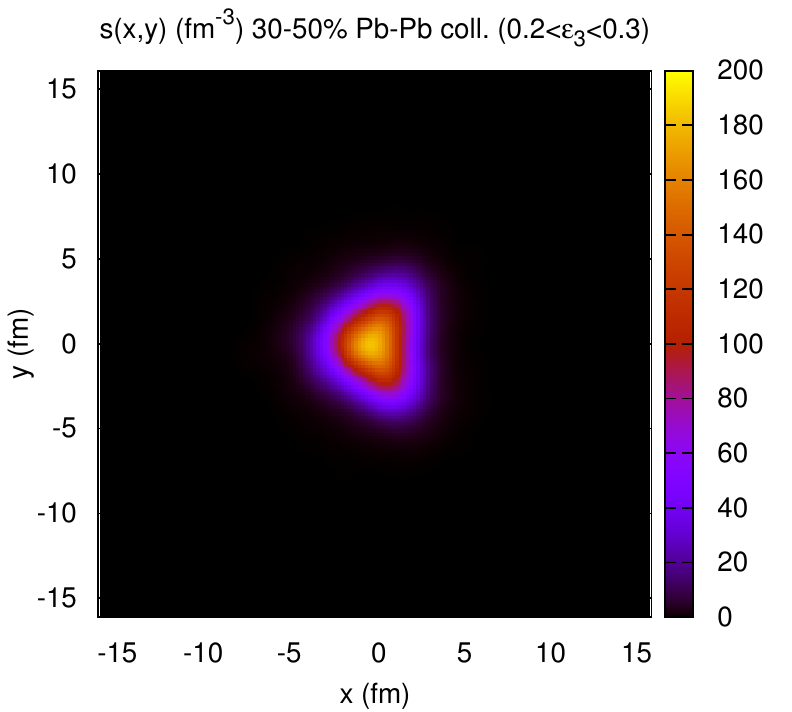}
\caption{The initial entropy-density for Pb-Pb collisions at $\sqrt{s_{\rm NN}}\!=\!5.02$ TeV of fixed triangular eccentricity $0.2\le\epsilon_3\le 0.3$ and different centrality classes.}\label{fig:incond-e3fixed} 
\end{center}
\end{figure}
\begin{figure}[!ht]
\begin{center}
\includegraphics[clip,width=0.48\textwidth]{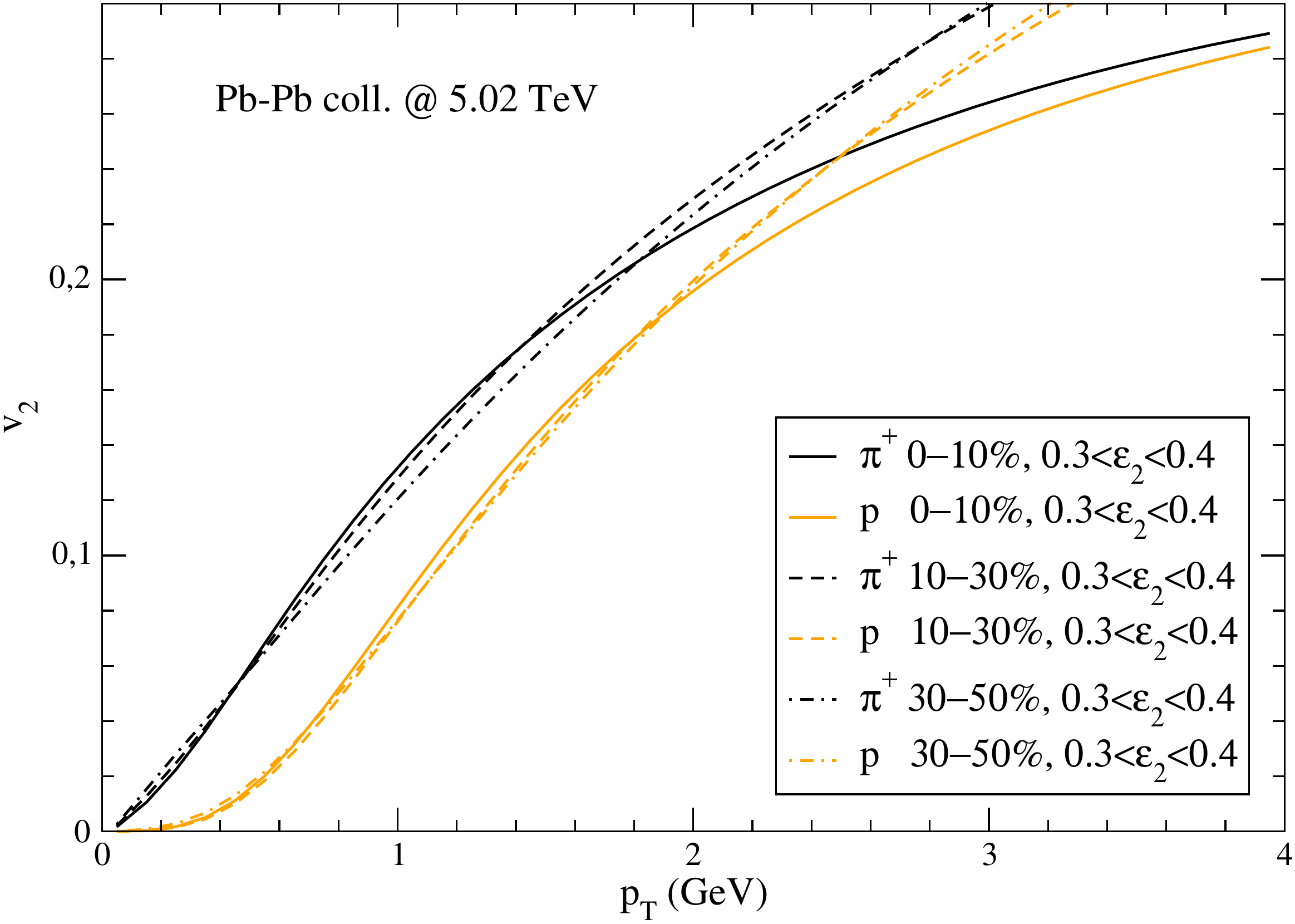}
\includegraphics[clip,width=0.48\textwidth]{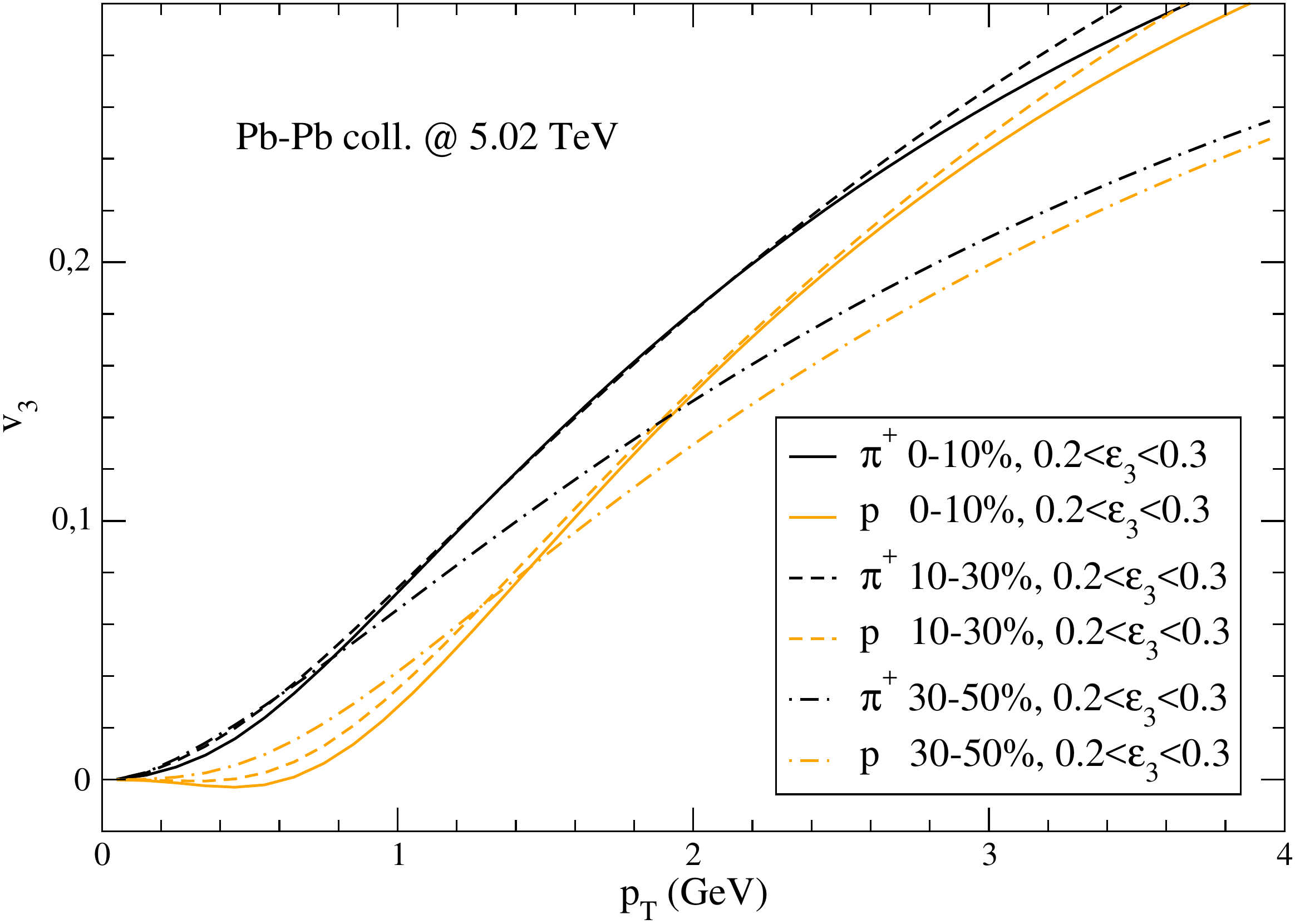}
\caption{The elliptic and triangular flow of pions and protons in Pb-Pb collisions at $\sqrt{s_{\rm NN}}\!=\!5.02$ TeV for events characterized by the same eccentricity ($0.3\le\epsilon_2\le 0.4$ and $0.2\le\epsilon_3\le 0.3$) but belonging to different centrality classes. At low $p_T$ the results are mainly sensitive to the eccentricity of the initial condition rather than to the centrality class.}\label{fig:vn_light-validation-fixed} 
\end{center}
\end{figure}
Going back to Fig.~\ref{fig:dNdeps} we note how the eccentricity distributions of different centrality classes display a significant overlap: we can have events in which the system is initially characterized by an equal degree of geometric deformation, but by very different dimensions and energy density. As a complementary study we select, in the different centrality classes, samples of events with an initial geometric asymmetry belonging to the same narrow interval: we choose $0.3\le\epsilon_2\le 0.4$ and $0.2\le\epsilon_3\le 0.3$ for the elliptic and triangular deformation. The resulting average initial conditions of the systems are displayed in Figs.~\ref{fig:incond-e2fixed} and~\ref{fig:incond-e3fixed} for the 0-10\%, 10-30\% and 30-50\% centrality classes. Starting from such initial state it is then of interest to study the flow of light hadrons decoupling from the medium at the end of its hydrodynamic evolution, in order to check whether the angular distributions of final-state particles respond only to the initial geometry of the system or whether its very different energy density plays a role. As one can see from Fig.~\ref{fig:vn_light-validation-fixed} the major role is played by the initial geometric deformation: the $v_n$ curves for pions and protons in different centrality classes display a strong overlap for a quite extended range of transverse momentum $p_T$ (with the partial exception of the triangular flow for the 30-50\% class). It is of interest to perform the same study for the case of heavy-flavour particles, since their energy-loss and degree of thermalization should be sensitive to the dimension and transport coefficients of the medium, both depending strongly on the centrality of the collision.

\section{Heavy flavour transport and hadronization}\label{Sec:transport}
After modelling the initial state with the Glauber-MC approach described in Sec.~\ref{Sec:medium}, obtaining an average initial condition for the selected subsample of collisions (with cuts on centrality and eccentricity), heavy quarks are distributed in the transverse plane according to the local density of binary collisions. Their propagation in the medium is then studied through the relativistic Langevin equation
\beq
{\Delta \vec{p}}/{\Delta t}=-{\eta_D(p)\vec{p}}+{\vec\xi(t)}.\label{eq:Langevin}
\eeq
containing a deterministic friction force quantified by the drag coefficient $\eta_D$ and a random noise term specified by its temporal correlator
\beq
\langle\xi^i(\vec p_t)\xi^j(\vec p_{t'})\rangle\!=\!{b^{ij}(\vec p_t)}{\delta_{tt'}}/{\Delta t}\qquad{b^{ij}(\vec p)}\!\equiv\!{\kk_\|(p)}\pp^i\pp^j+{\kk_\perp(p)}(\delta^{ij}\!-\!\pp^i\pp^j).
\eeq
In the above, information on the background medium provided by hydrodynamics enters in two ways: first through its collective velocity flow $u^\mu$, whose knowledge is necessary in order to perform the update of the heavy-quark momentum at each time-step in the local rest frame of the fluid; secondly through the temperature dependence of the transport coefficients $\kappa_{\perp/\|}$ and $\eta_D$, which quantify the coupling of the heavy quarks with the medium. Our simulations are performed adopting two independent choices for the above transport coefficients, from weak-coupling (Hard-Thermal Loop~\cite{Alberico:2013bza}) and lattice-QCD calculations~\cite{Banerjee:2011ra,Francis:2015daa}. The differences in the final particle distributions obtained with these two sets of transport coefficients provide an estimate of the current theoretical uncertainties and of the potential discriminating power of the experimental data.
Lattice-QCD calculations provide in principle a non-perturbative result. However, they refer to the case of static, infinitely heavy, quarks; furthermore so far they are limited to the case of a pure gluon plasma and are affected by the unavoidable systematic uncertainties in extracting real-time information from simulations performed in an Euclidean spacetime. On the other hand weak-coupling calculations can deal with the realistic case of finite-mass quarks with relativistic momenta, but they are so far limited at the tree-level, with resummation of medium effects in the gluon propagators.

Hadronization is modelled recombining at freeze-out the heavy quarks with light thermal partons from the same fluid-cell (an instantaneous decoupling with no further rescattering in the hadron-gas phase is assumed), forming $Q\overline q$ (or $\overline Q q$) strings which are then fragmented according to the Lund model implemented in PYTHIA 6.4. This turns out to have a major effect on the final charm and beauty hadron distributions. 
For deeper details about the implementation of our transport calculations and our modelling of hadronization we refer the reader to our past publications~\cite{Beraudo:2014boa,Beraudo:2017gxw}. For a comprehensive review of transport calculations applied to the study of heavy-flavour observables in relativistic heavy-ion collisions, emphasizing the various source of systematic uncertainties and the state of the art of the extraction of transport coefficients, see for instance Refs.~\cite{Rapp:2018qla,Cao:2018ews}.

\section{Results}\label{Sec:results}
\begin{figure}[!ht]
\begin{center}
  \includegraphics[clip,width=0.8\textwidth]{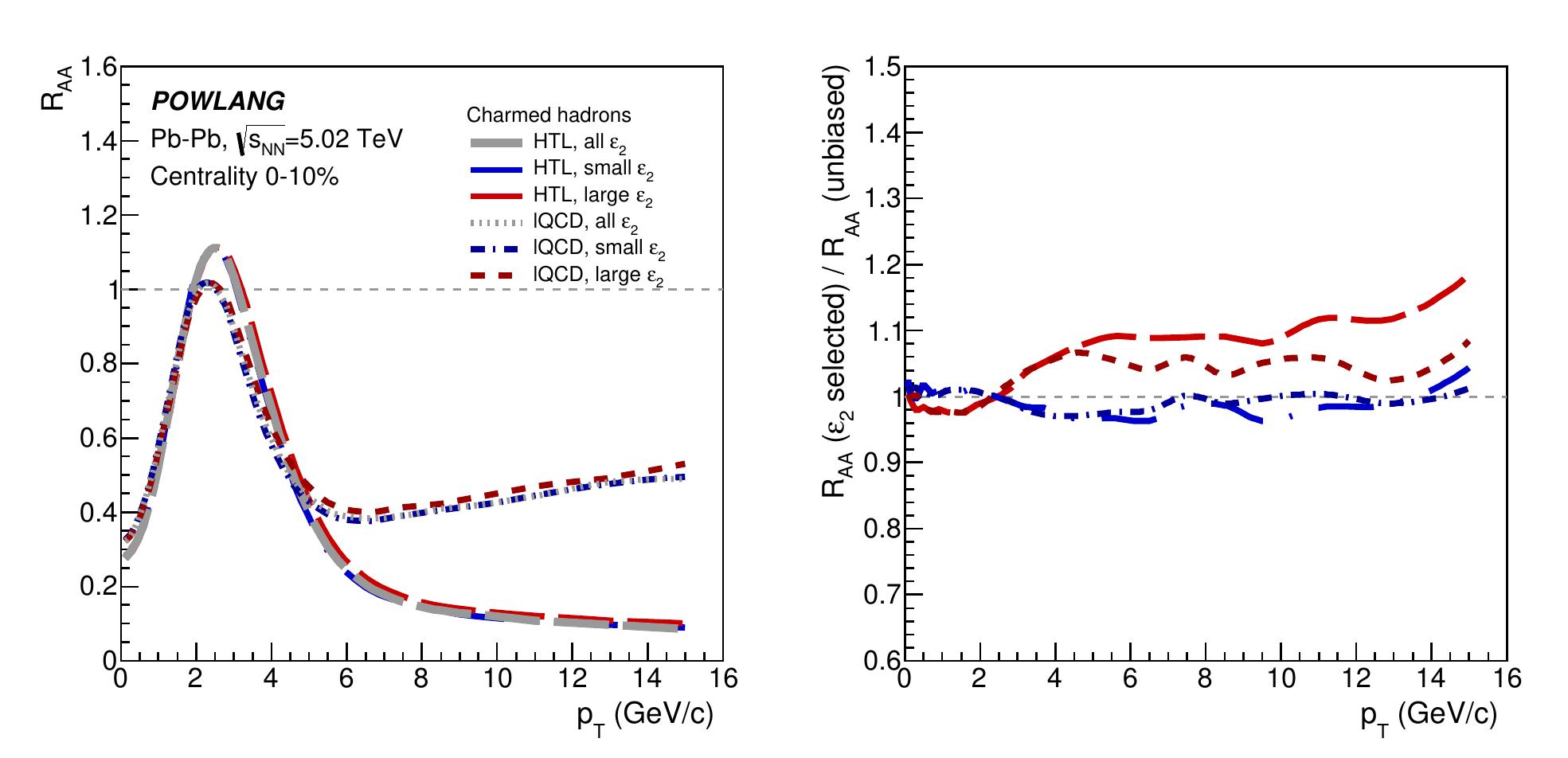}
  \includegraphics[clip,width=0.8\textwidth]{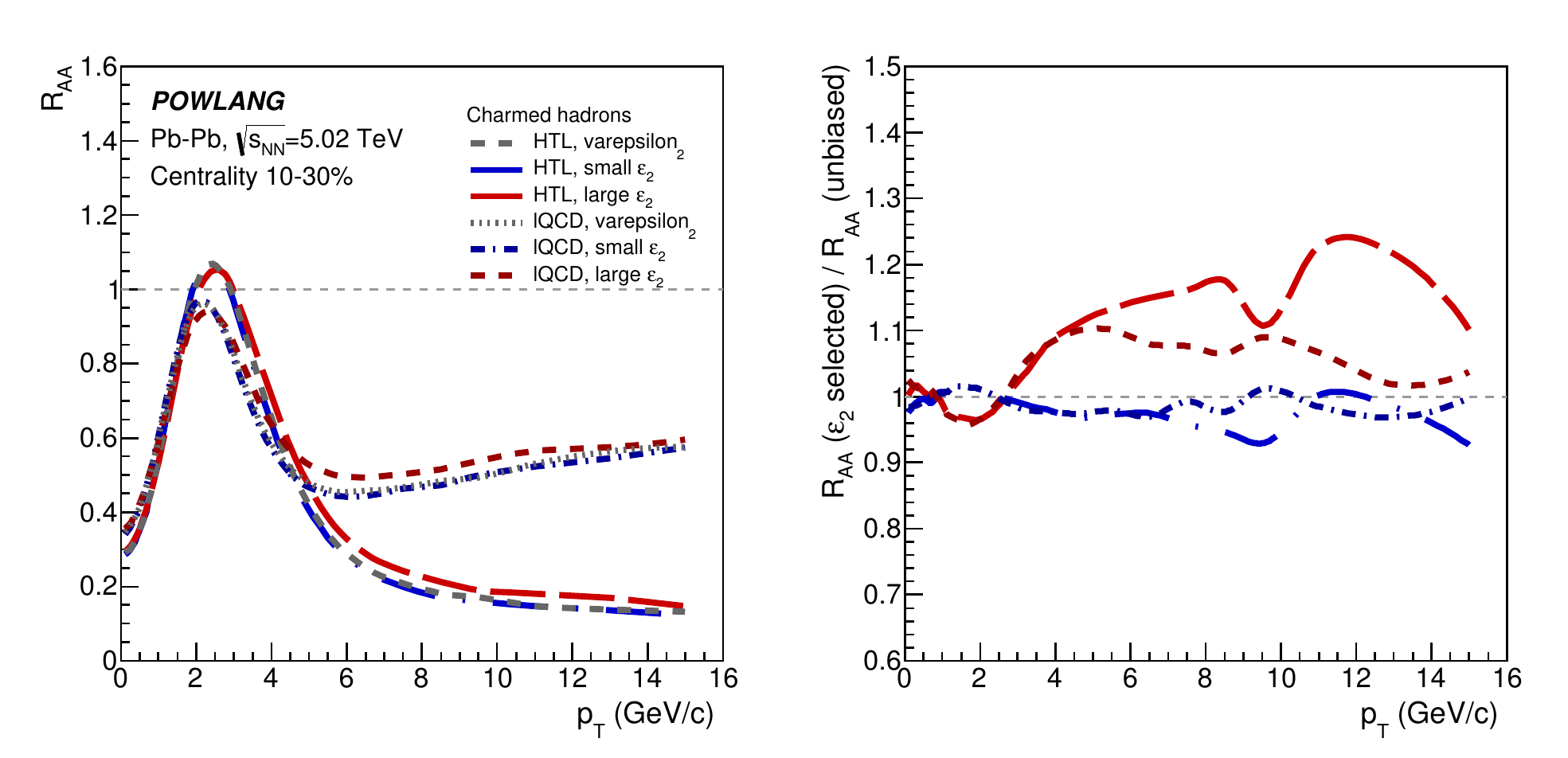}
\includegraphics[clip,width=0.8\textwidth]{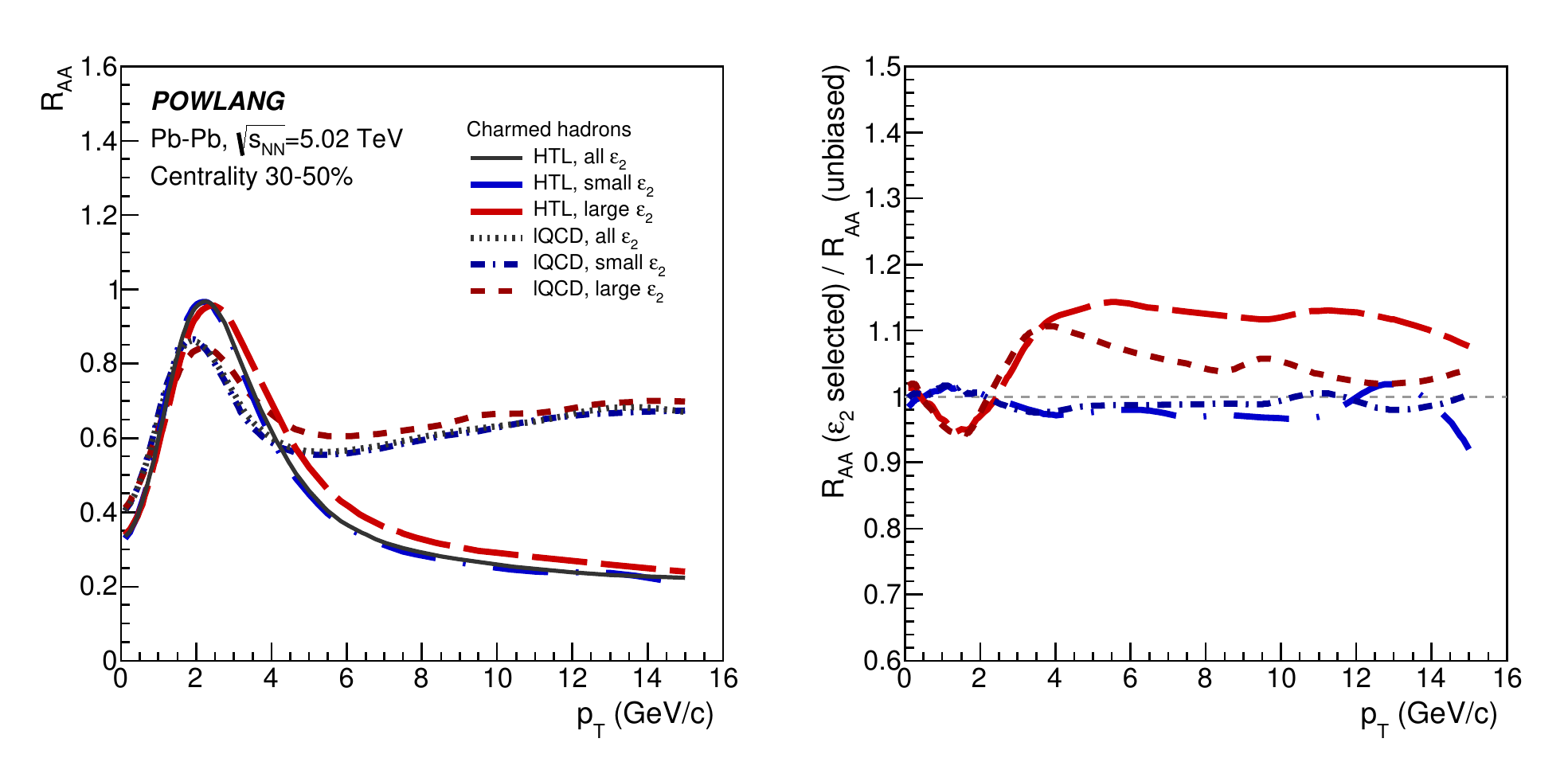}  
\caption{The nuclear modification factor $R_{\rm AA}$ of charmed hadrons in the 0-10\% (top), 10-30\% (middle) and 30-50\% (bottom) most central Pb-Pb collisions at $\sqrt{s_{\rm NN}}\!=\!5.02$ TeV. Results for the 20\% highest-$\epsilon_2$ and the 60\% lowest-$\epsilon_2$ selection of events are compared to the unbiased case. For both choices of transport coefficients the results display only a mild and similar sensitivity to the initial eccentricity.}\label{fig:RAAQ_sel-vs-mb-all} 
\end{center}
\end{figure}

We start our event-shape-engineering study of heavy-flavour production in nucleus-nucleus collisions considering the nuclear modification factor of charmed hadrons. We consider the 0-10\%, 10-30\% and 30-50\% centrality classes and we compare the results obtained for subsamples of events corresponding -- for a given centrality -- to an unbiased selection, to the 20\% highest $\epsilon_2$ and to the 60\% lowest $\epsilon_2$. As one can see from Fig.~\ref{fig:RAAQ_sel-vs-mb-all}, the effect of the eccentricity cuts is quite modest, at most of the order 10-20\% when considering the high-$\epsilon_2$ sample, and consistent with the anticorrelation between eccentricity and centrality: on average, within a given centrality class, high-$\epsilon_2$ events correspond also to a larger impact parameter and hence to a lower initial size and density of the system. We remind that, in order to remove this bias and get a cleaner decoupling between the density and the elliptic asymmetry of the medium, the ALICE collaboration performed the selection on eccentricity in very narrow centrality intervals, corresponding to bins of 1\% of the total hadronic cross-section. This was possible taking advantage of the large available statistics: for each centrality class considered in the experimental analysis (10-30\% and 30-50\%) a number of events of order $10^7$ was collected. For this first theroretical study on the subject we rely on a less demanding approach in terms of computing and storage resources, bearing in mind in interpreting our findings the not complete decoupling between the system density/size and its geometrical deformation in our selection of events. Notice that, even considering very narrow centrality bins as done by the ALICE collaboration in Ref.~\cite{Adam:2015eta}, a selection on centrality can lead to an effect on the transverse-momentum distributions: Glauber-MC simulations show a positive correlations between the initial density of the system and its eccentricity and this could explain the larger radial flow of light hadrons observed in events with larger average elliptic flow.

\begin{figure}[!ht]
\begin{center}
  \includegraphics[clip,width=0.8\textwidth]{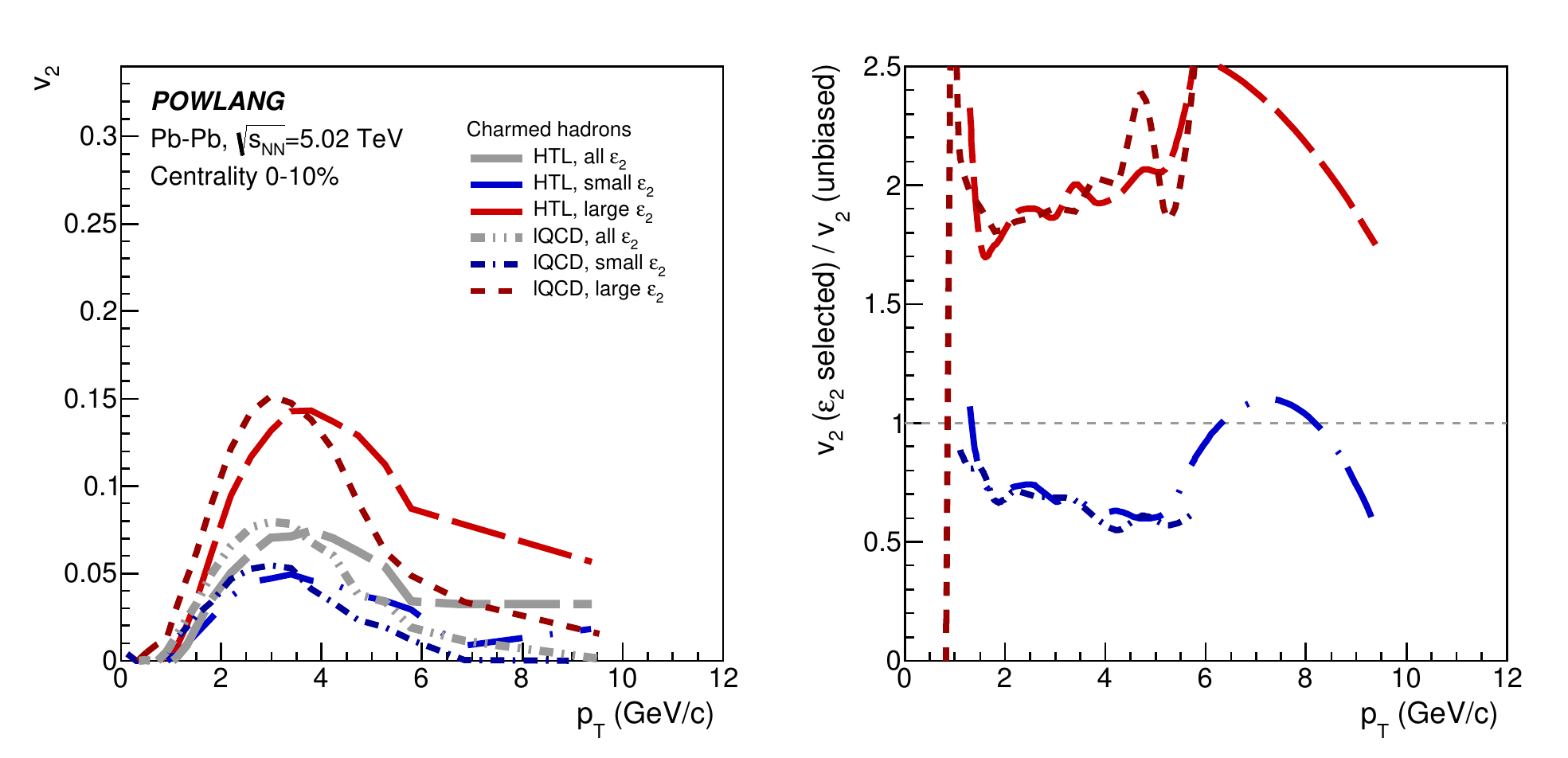}
  \includegraphics[clip,width=0.8\textwidth]{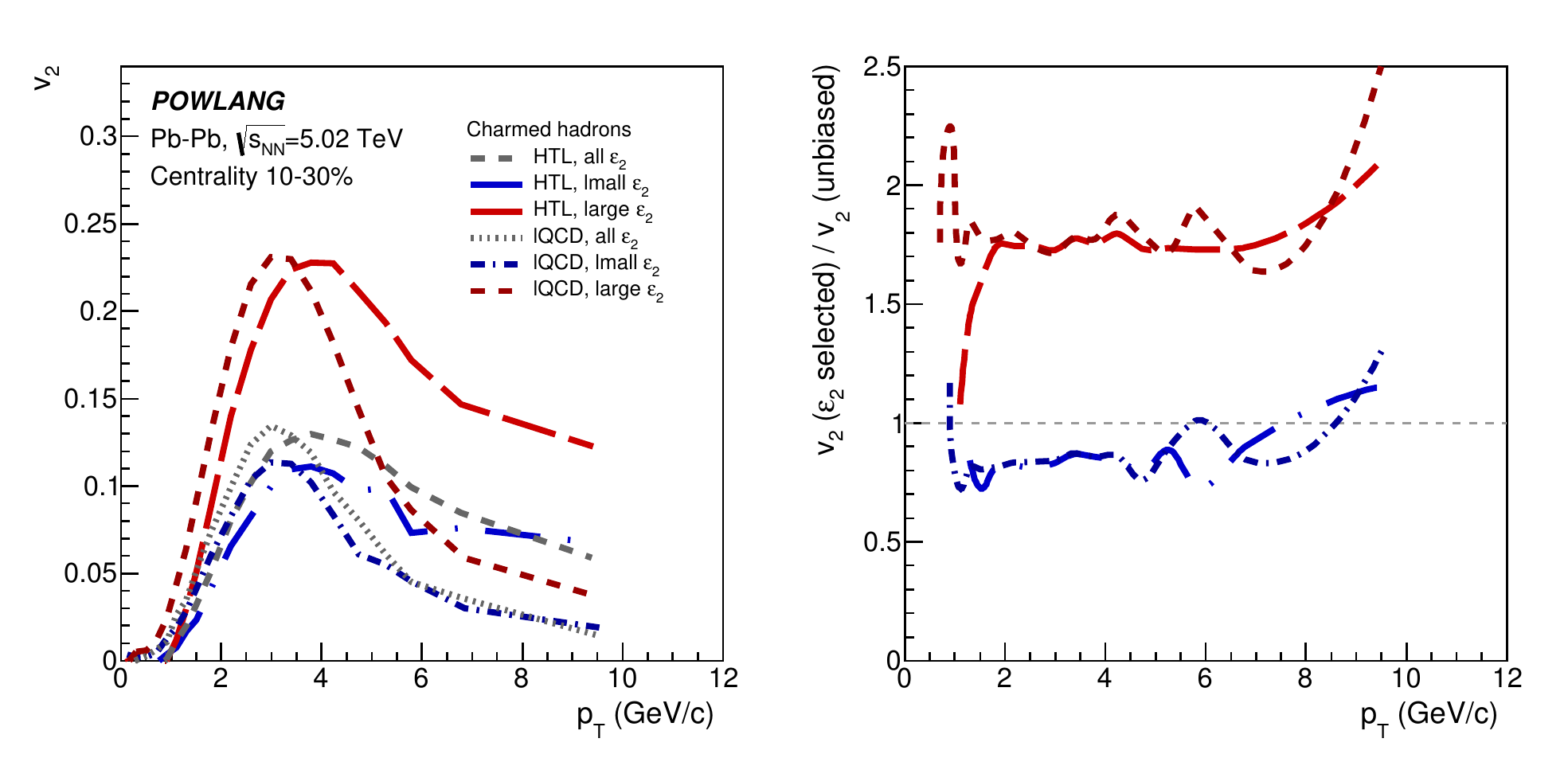}
\includegraphics[clip,width=0.8\textwidth]{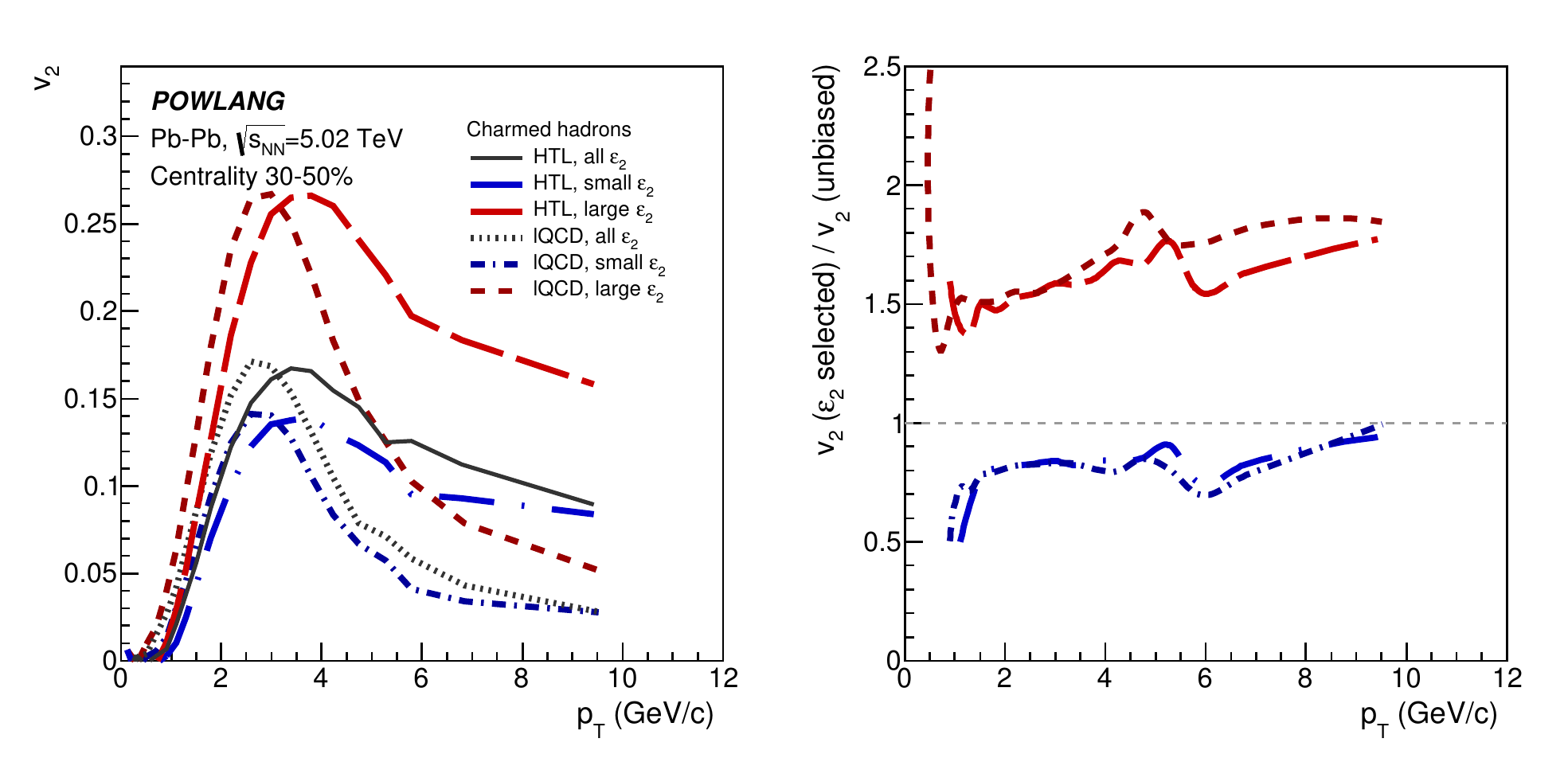}  
\caption{The elliptic-flow coefficient $v_2$ of charmed hadrons in the 0-10\% (top), 10-30\% (middle) and 30-50\% (bottom) most central Pb-Pb collisions at $\sqrt{s_{\rm NN}}\!=\!5.02$ TeV. Results for the 20\% highest-$\epsilon_2$ and to the 60\% lowest-$\epsilon_2$ selection of events are compared to the unbiased case. For both choices of transport coefficients the results display a strong and similar sensitivity to the initial eccentricity.}\label{fig:v2Q_sel-vs-mb-all} 
\end{center}
\end{figure}
We now move to the study of the elliptic flow. As one can see in Fig.~\ref{fig:v2Q_sel-vs-mb-all}, referring to the 0-10\%, 10-30\% and 30-50\% centrality classes, the effect of event-shape engineering is much larger in this case. If we focus on the ratio $v_2^{{\rm high}\!-\!\epsilon_2}/v_2^{\rm unbiased}$, for all the three centrality classes the 20\% highest-$\epsilon_2$ events display an average $v_2$ coefficient almost twice as large as the one found in the unbiased sample. The size of the effect looks quite independent of the transport coefficients and the transverse momentum of the charmed hadron. Similar considerations hold for the ratio $v_2^{{\rm low}-\epsilon_2}/v_2^{\rm unbiased}$, which looks quite flat around 0.7-0.8 for a sufficiently broad range $p_T$ and independent of the coupling of the heavy quark with the medium (HTL vs lQCD curves in the figures). Actually, comparing the various centralities, the effect on $v_2$ of selecting high-eccentricity events looks larger in the 0-10\% class, in agreement with what already found for pions and displayed in the left panel of Fig.~\ref{fig:v2vsv2mb}. To summarize, the charm-hadron $v_2$ calculated with different transport coefficients displays non-negligible differences, in particular for $p_T\gsim 4$ GeV/c, but its relative variation once selecting subsample of events with higher/lower eccentricity looks very similar.

\begin{figure}[!ht]
\begin{center}
\includegraphics[clip,width=0.8\textwidth]{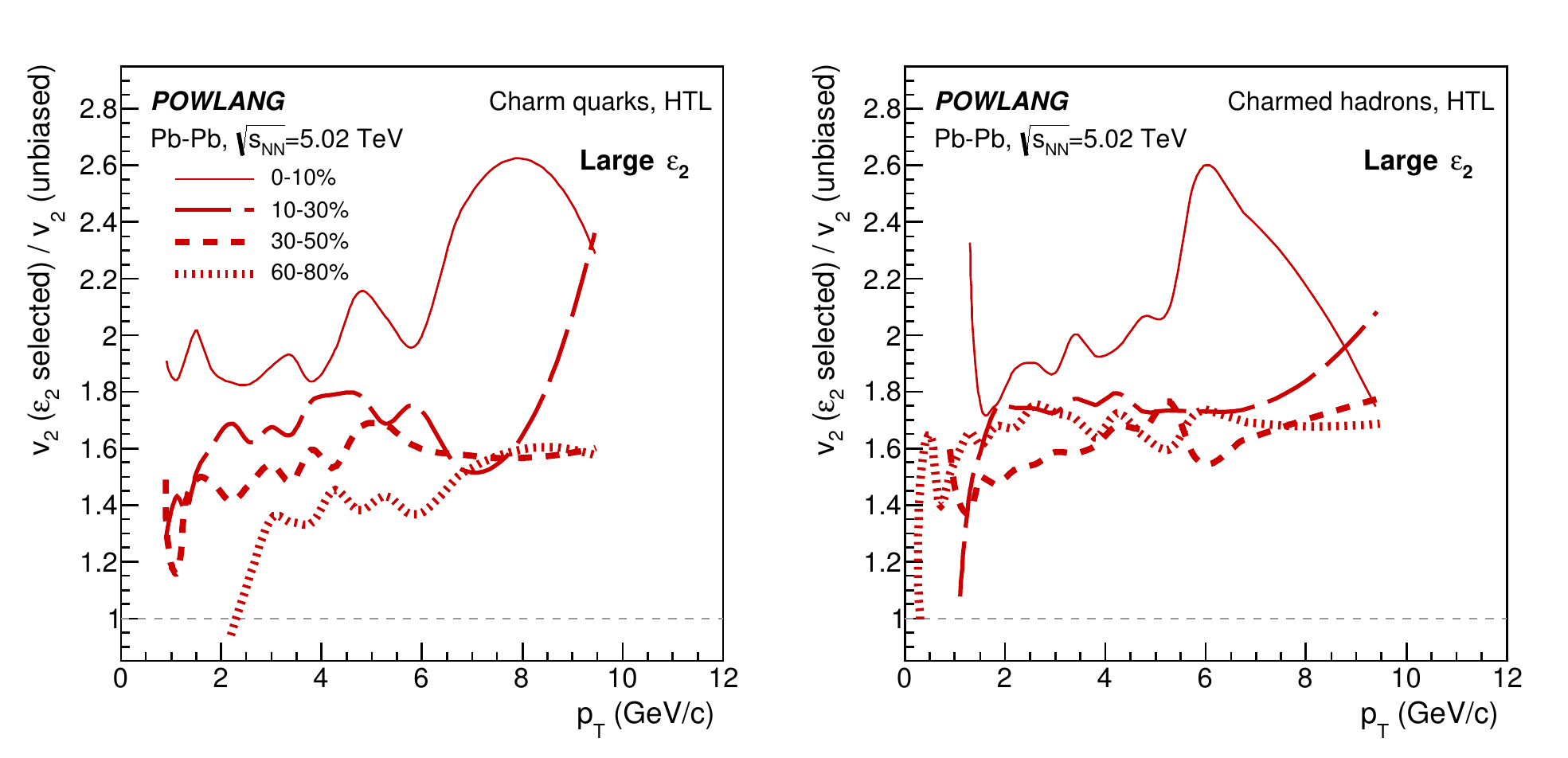}
\caption{The ratio of the $v_2$ coefficients for charm (quarks in the left panel, hadrons in the right panel) in high-eccentricity over unbiased events, for different centrality classes.}\label{fig:v2Q_high-vs-mb} 
\end{center}
\end{figure}
In order to better assess how the effect of the eccentricity selection varies with the centrality of the collision in Fig.~\ref{fig:v2Q_high-vs-mb} we plot the ratio $v_2^{{\rm high}-\epsilon_2}/v_2^{\rm unbiased}$ for charm quarks and hadrons in various classes of $N_{\rm coll}$. Here we include also a very peripheral (60-80\%) sample of events. As one can see in the left panel, curves at the quark level display a clear ordering in centrality. The enhancement of the charm-quark $v_2$ when selecting high-eccentricity events gets larger moving from peripheral to central collisions: the denser and larger the medium, the stronger its effect on the propagation of charm quarks and hence the more evident the signatures of its asymmetric geometry and flow in the final particle distributions. Actually, varying the centrality seems to play a milder role at the hadron level: hadronization modelled via recombination with light partons from the medium probably washes out part of the effect. Also experimental data by ALICE~\cite{Acharya:2018bxo}, although affected by quite large error bars and dependent on the eccentricity estimator, seem to indicate that there is not a big dependence on centrality, at least for the classes considered in their analysis (10-30\% and 30-50\%).     

\begin{figure}[!ht]
\begin{center}
  \includegraphics[clip,width=0.8\textwidth]{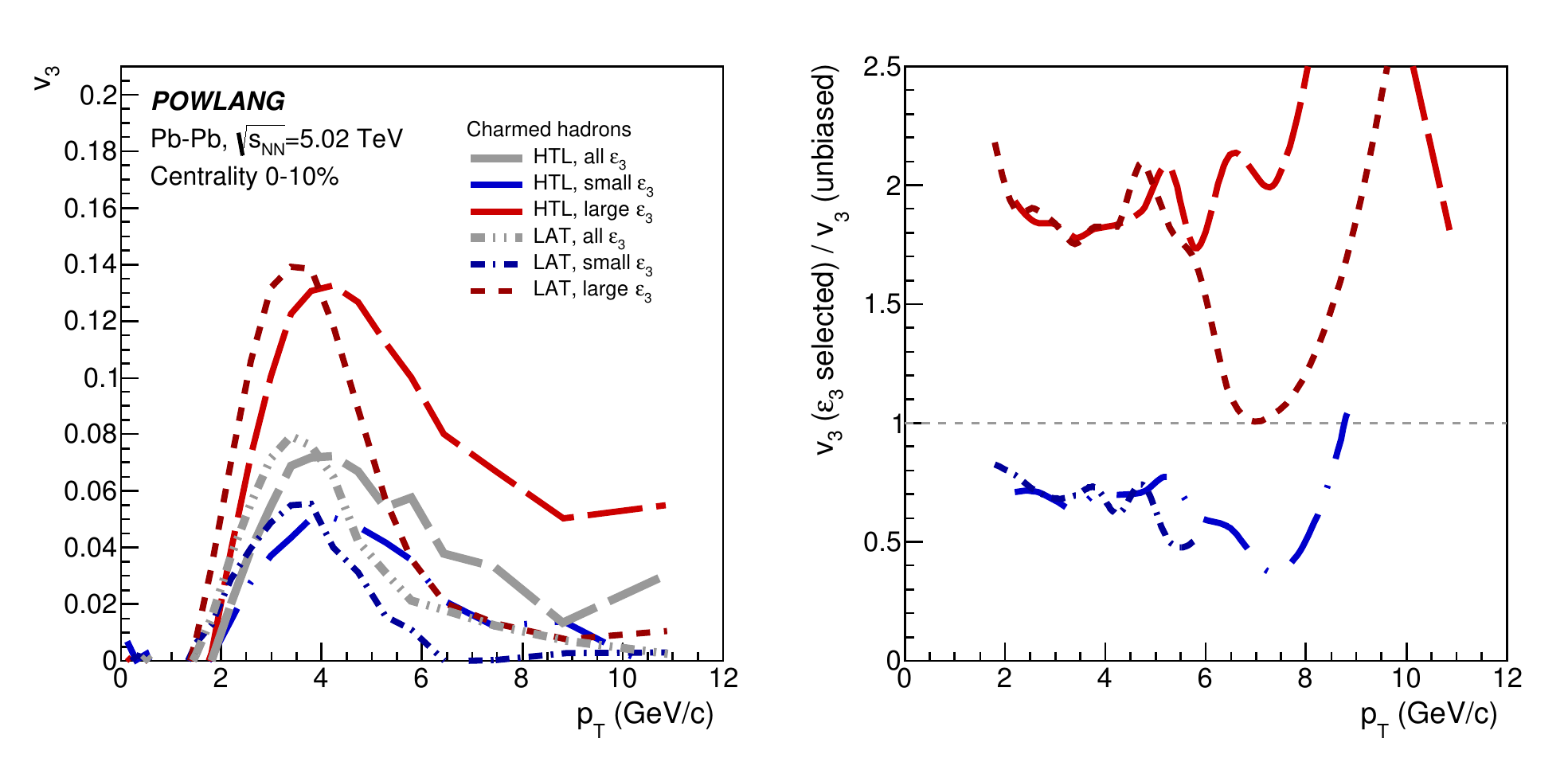}
  \includegraphics[clip,width=0.8\textwidth]{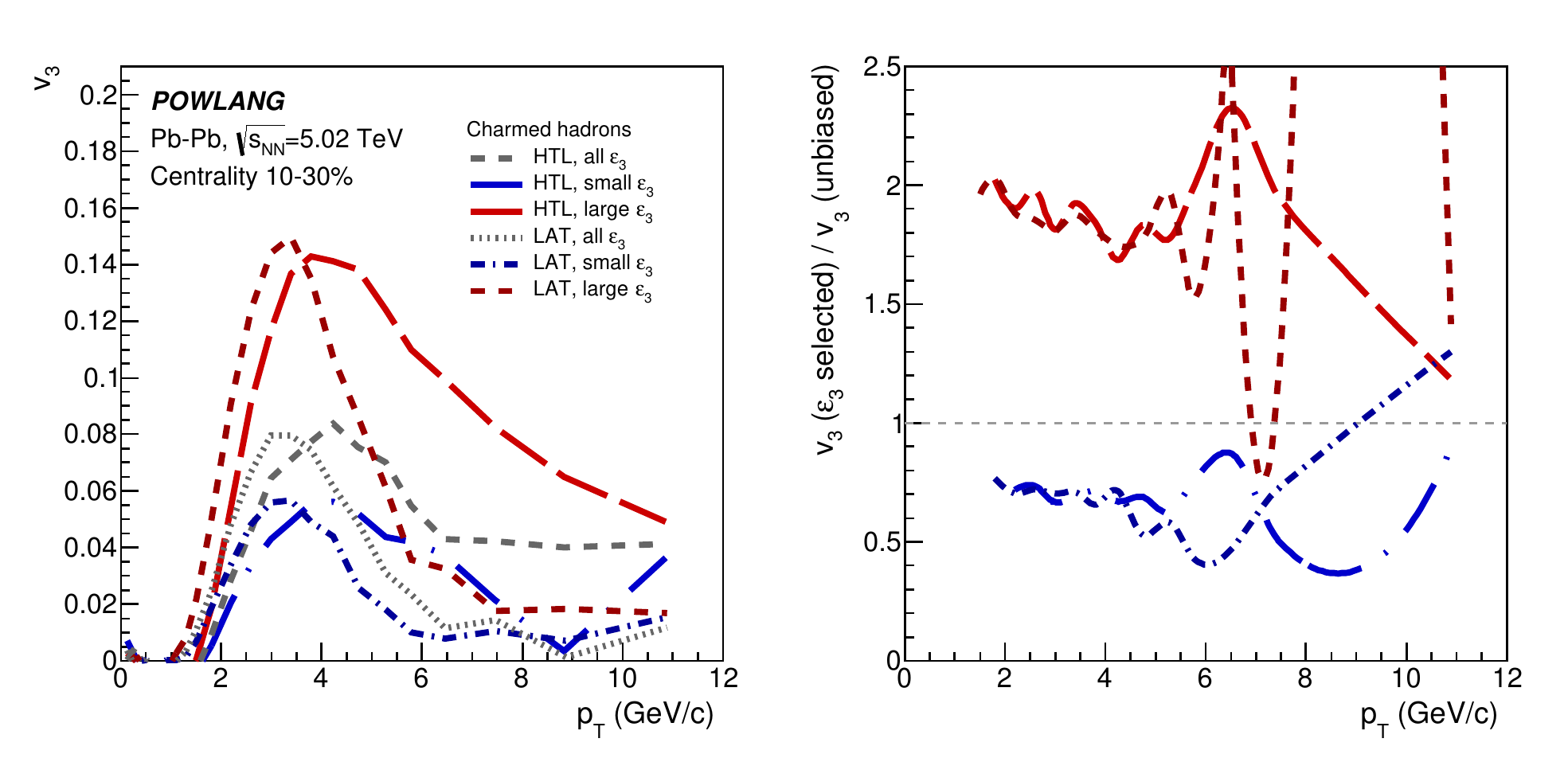}
\includegraphics[clip,width=0.8\textwidth]{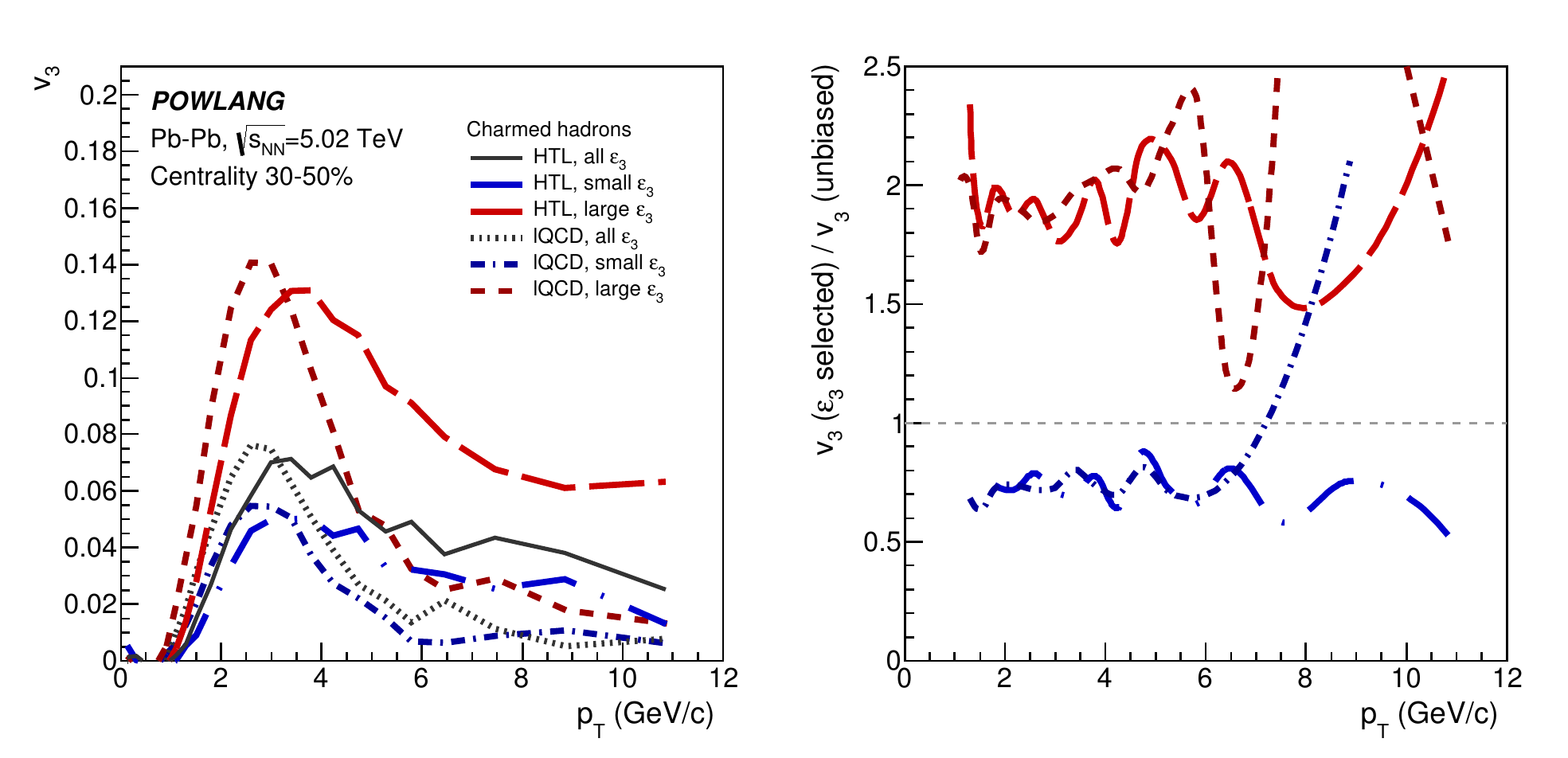}  
\caption{The triangular-flow coefficient $v_3$ of charmed hadrons in the 0-10\% (top), 10-30\% (middle) and 30-50\% (bottom) most central Pb-Pb collisions at $\sqrt{s_{\rm NN}}\!=\!5.02$ TeV. Results for the 20\% highest-$\epsilon_3$ and to the 60\% lowest-$\epsilon_3$ selection of events are compared to the unbiased case. For both choices of transport coefficients the results display a strong and similar sensitivity to the initial eccentricity.}\label{fig:v3Q_sel-vs-mb-all} 
\end{center}
\end{figure}
Although not yet considered in the experimental analysis, it is of interest to perform the same event-shape-engineering study for the triangular flow $v_3$. Remember that in this case, for any selection on centrality and eccentricity $\epsilon_3$, an average initial condition is built after rotating each event in the transverse plane by -$\Psi_3$, as described in Sec.~\ref{Sec:medium}. Our findings are shown in Fig.~\ref{fig:v3Q_sel-vs-mb-all} and look similar to the ones obtained for the elliptic flow. Considering the ratio $v_3^{{\rm high}\!-\!\epsilon_3}/v_3^{\rm unbiased}$, for all the three centrality classes the 20\% highest-$\epsilon_3$ events display an average $v_3$ coefficient almost twice as large as the one found in the unbiased sample. The size of the effect looks quite independent of the transport coefficients and the transverse momentum of the charmed hadron, although fluctuations look very large in $p_T$ regions in which the signal is small. The effect looks also pretty independent of the centrality of the collision, as already found for pions and shown in the right panel of Fig.~\ref{fig:v2vsv2mb}. Also in the case of the low-$\epsilon_3$ subsamples deviations from the unbiased results are of the same order of what found for the elliptic flow.

\begin{figure}[!ht]
\begin{center}
  \includegraphics[clip,width=0.48\textwidth]{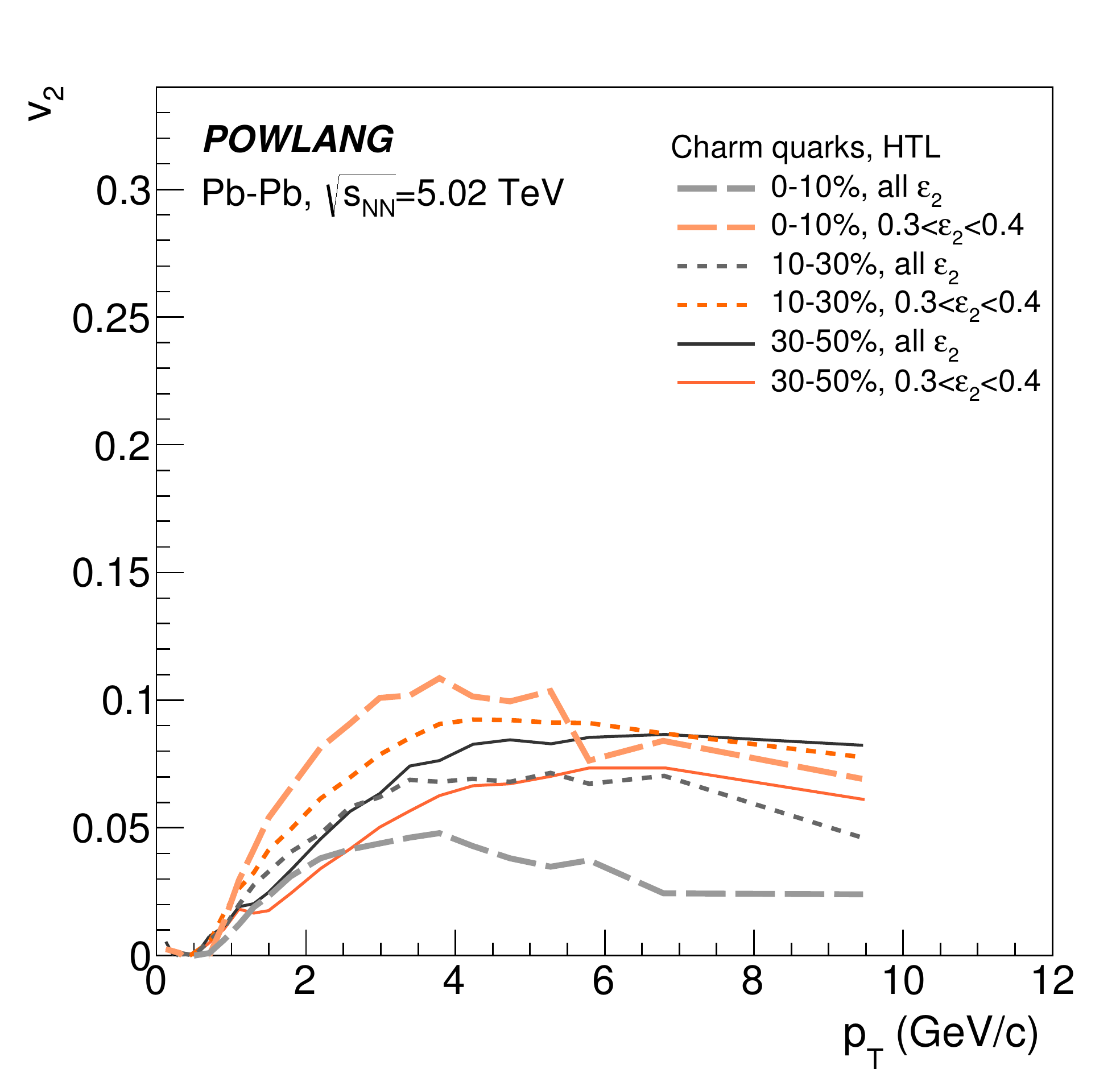}
  \includegraphics[clip,width=0.48\textwidth]{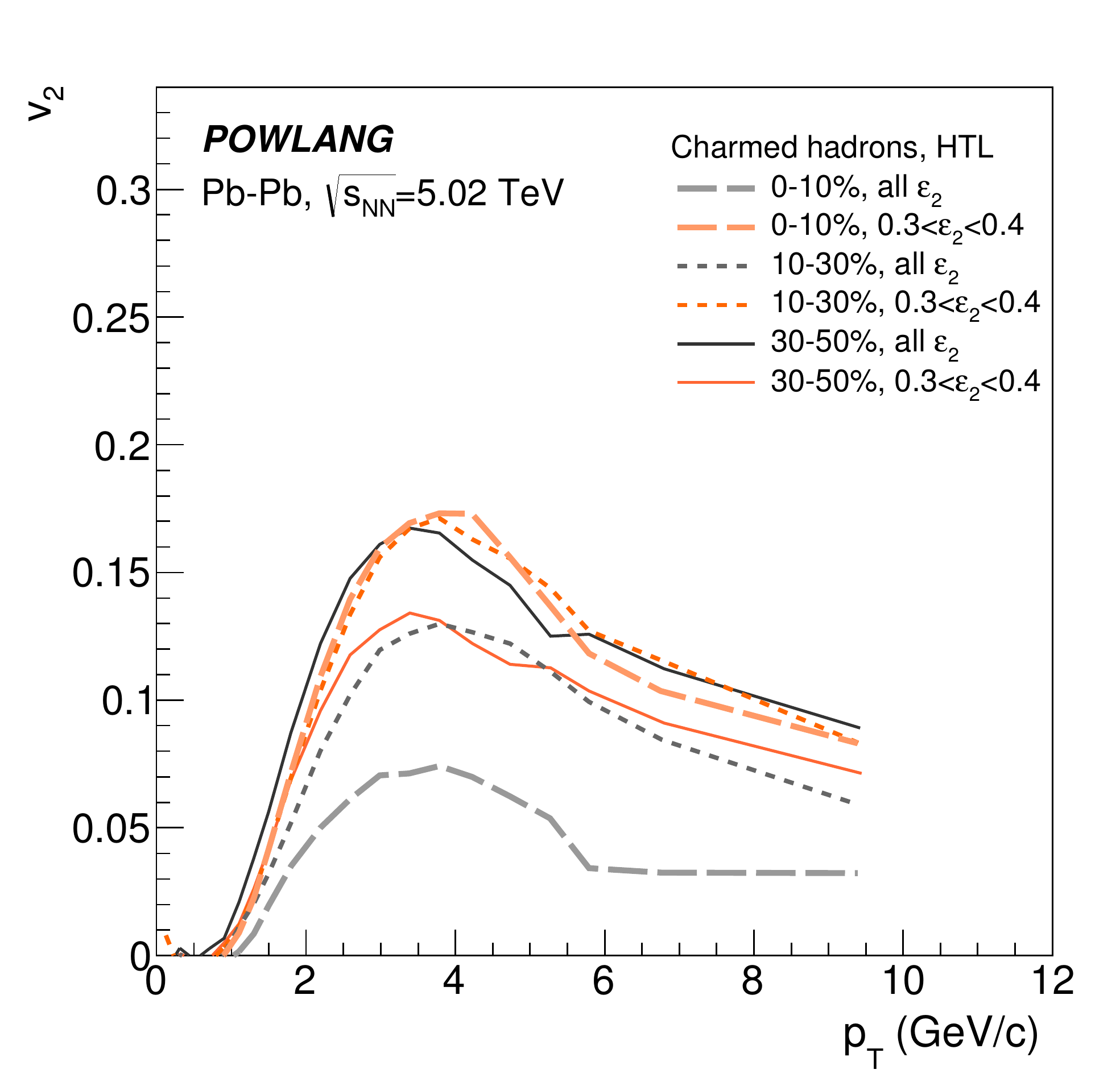}  
\caption{The elliptic-flow coefficient $v_2$ of charmed quarks (left panel) and hadrons (right panel) in Pb-Pb collisions at $\sqrt{s_{\rm NN}}\!=\!5.02$ TeV for various centrality classes. For each class we display results referring to the fixed eccentricity interval $0.3\le\epsilon_2\le 0.4$ and to an unbiased selection of the events. At the quark level results for events with the same $\epsilon_2$ but belonging to different centrality classes display sizable differences, the response to the initial eccentricity being stronger for more central events. The difference is partially washed-out by hadronization via recombination.}\label{fig:v2_e2-fixed} 
\end{center}
\end{figure}
\begin{figure}[!ht]
\begin{center}
  \includegraphics[clip,width=0.48\textwidth]{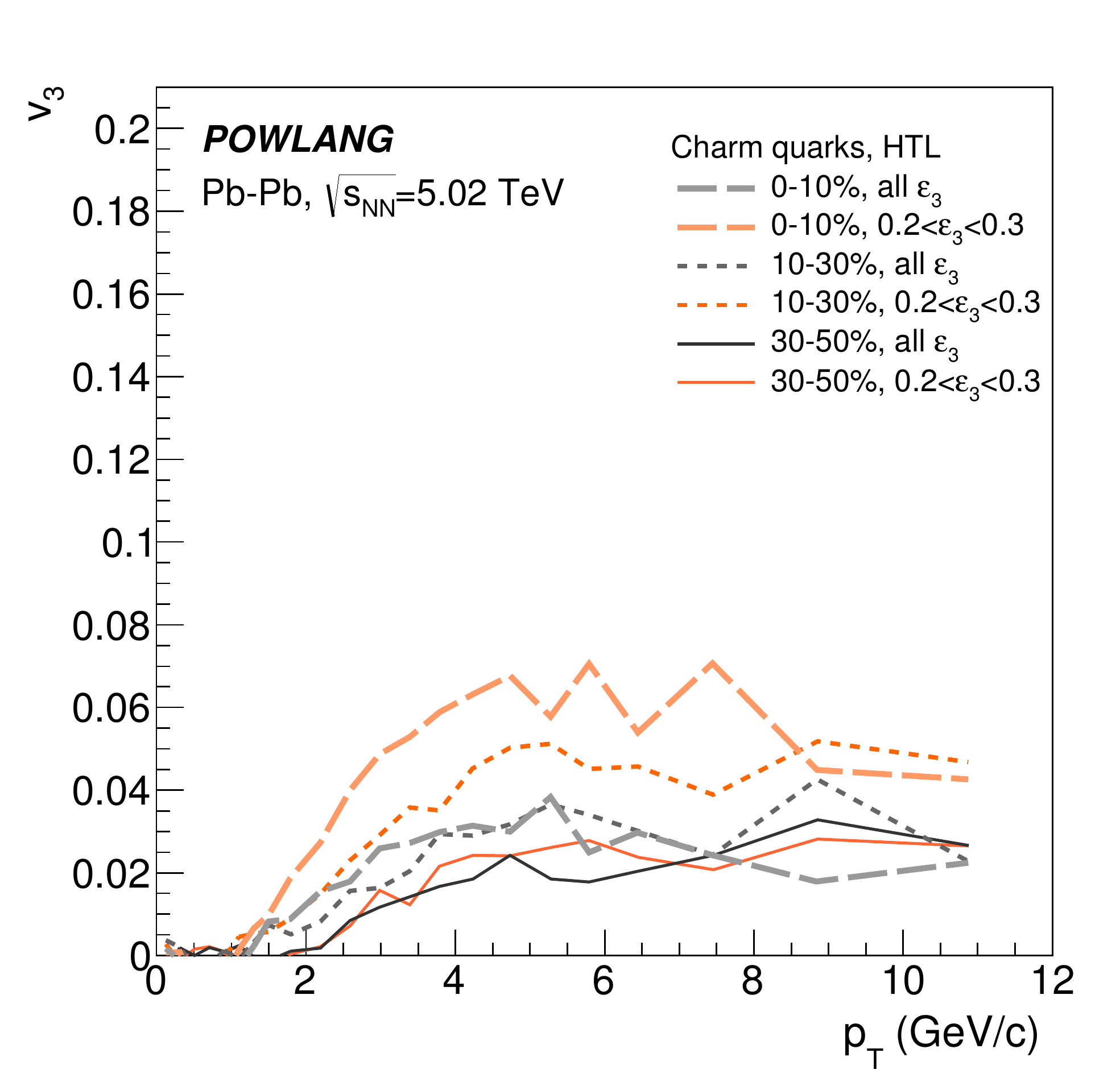}
  \includegraphics[clip,width=0.48\textwidth]{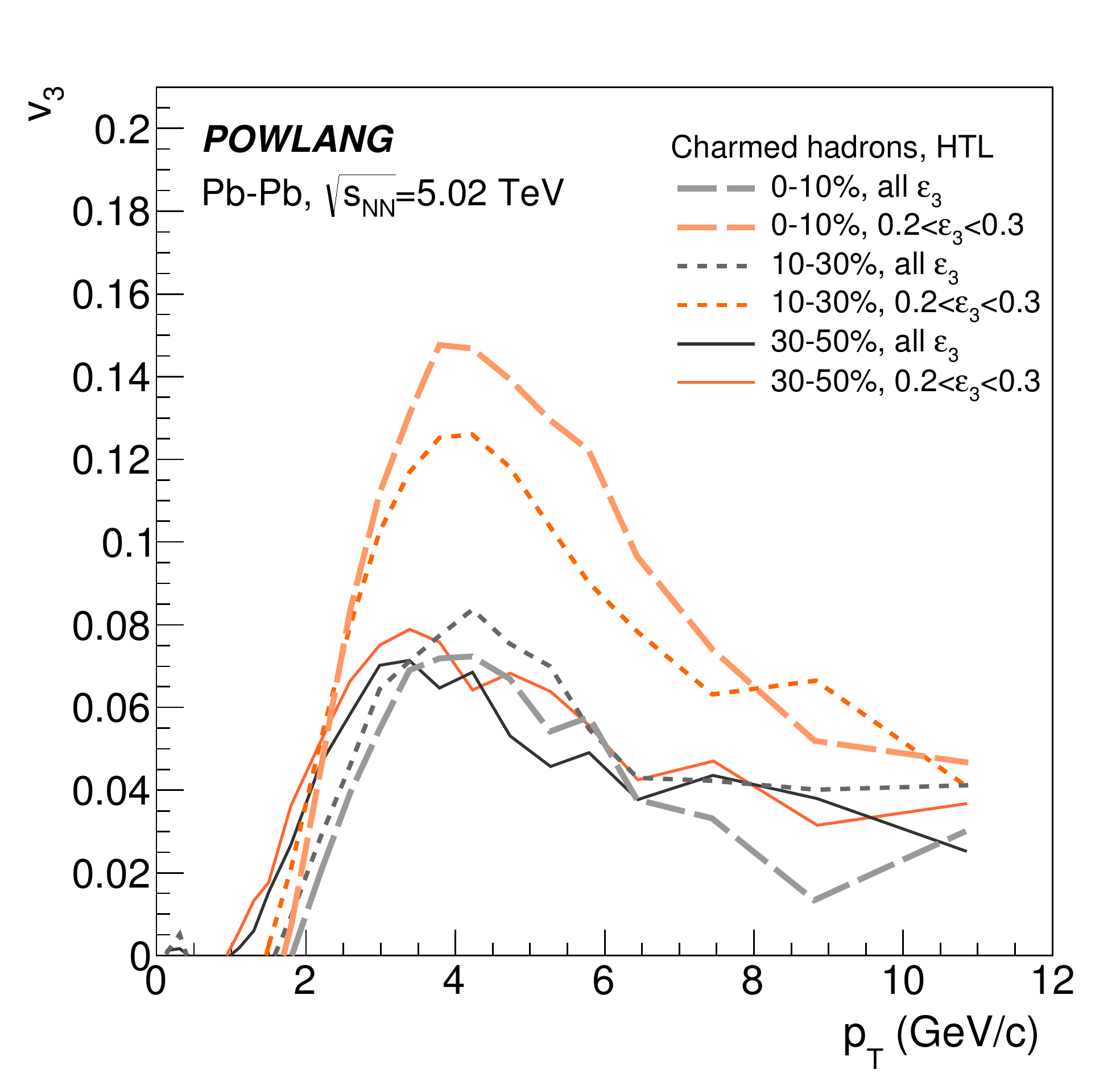}  
\caption{The triangular-flow coefficient $v_3$ of charmed quarks (left panel) and hadrons (right panel) in Pb-Pb collisions at $\sqrt{s_{\rm NN}}\!=\!5.02$ TeV for various centrality classes. For each class we display results referring to the fixed eccentricity interval $0.2\le\epsilon_3\le 0.3$ and to an unbiased selection of the events. At the quark level results for events with the same $\epsilon_3$ but belonging to different centrality classes display sizable differences, the response to the initial eccentricity being stronger for more central events. The difference persists even after hadronization via recombination.}\label{fig:v3_e3-fixed} 
\end{center}
\end{figure}
We decide now to adopt a different perspective and compare the results for the flow of heavy-flavour particles in events characterized by the same initial eccentricity, but belonging to different centrality classes. Results for the elliptic and triangular flow are displayed in Figs.~\ref{fig:v2_e2-fixed} and~\ref{fig:v3_e3-fixed}, referring to subsets of events with an initial asymmetry $0.3\le\epsilon_2\le 0.4$ and $0.2\le\epsilon_3\le 0.3$, respectively. Our scope is to point out differences among the results in the different centrality classes which can be attributed to the energy-density and size of the medium. In Sec.~\ref{Sec:medium}, in fact, we showed that the elliptic and triangular flow of light hadrons (pions and protons) decoupling from the fireball depends essentially on the initial eccentricity of the medium and only marginally on the centrality class (see Fig.~\ref{fig:vn_light-validation}): in spite of the very different temperature, size and lifetime of the medium, the final $v_2$ and $v_3$ of soft hadrons produced at hadronization look very similar. However the situation could be in principle different for heavy flavour particles, which do not come from the hadronization of the bulk medium itself, but whose parents are the $Q\overline Q$ pairs produced in hard scattering processes occurring before the formation of a thermalized quark-gluon plasma. These heavy quarks, before decoupling, interact strongly with the fireball through which they propagate and we expect that the different medium size, lifetime and temperature in the different centrality classes should affect the final results. This is clearly visible in the left panels of Figs.~\ref{fig:v2_e2-fixed} and~\ref{fig:v3_e3-fixed}: one gets very different results for the elliptic and triangular flow of charm quarks in events with the same initial eccentricity (orange curves) but belonging to different centrality classes, due to the different amount of energy-loss and diffusion suffered in the medium. Is the effect observable also in the final hadron distributions? As one can see from the right panels of Figs.~\ref{fig:v2_e2-fixed} and~\ref{fig:v3_e3-fixed} at the level of charmed hadrons deviations of the results among the different centrality classes are milder. This is particularly evident in the case of the elliptic flow. The curves for the $v_2$ corresponding to the unbiased selection of events (grey curves) look very different going from central to more peripheral collisions; on the contrary if we focus on events of various centrality but corresponding to a very similar initial eccentricity (orange curves) the curves tend to merge, although this was not the case at the quark level. This is clearly a consequence of hadronization, which in our model proceeds via recombination of the heavy quarks with the light thermal partons from the medium, characterized by a very similar anisotropic flow in the different centrality classes if one consider events of comparable initial eccentricity. Notice that a difference among events with the same eccentricity but belonging to different centrality classes persists in the case of the triangular flow of charm hadrons, as one can see in particular comparing the 30-50\% curve with the ones of the 0-10\% and 10-30\% centrality classes: this should not surprise us too much, since the same different was present also in the case of light hadrons (see Fig.~\ref{fig:vn_light-validation-fixed}).

\begin{figure}[!ht]
\begin{center}
  \includegraphics[clip,width=0.8\textwidth]{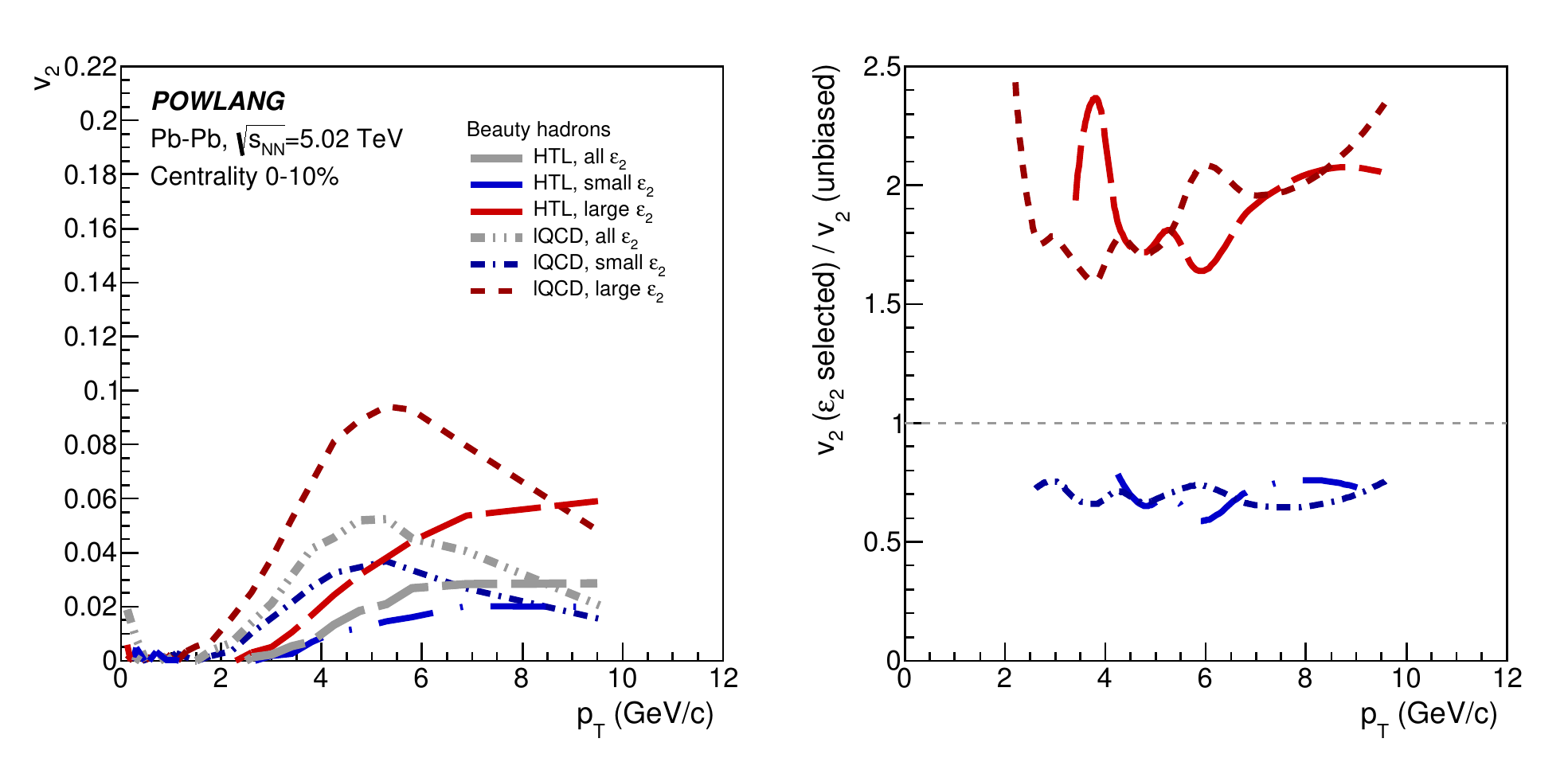}
  \includegraphics[clip,width=0.8\textwidth]{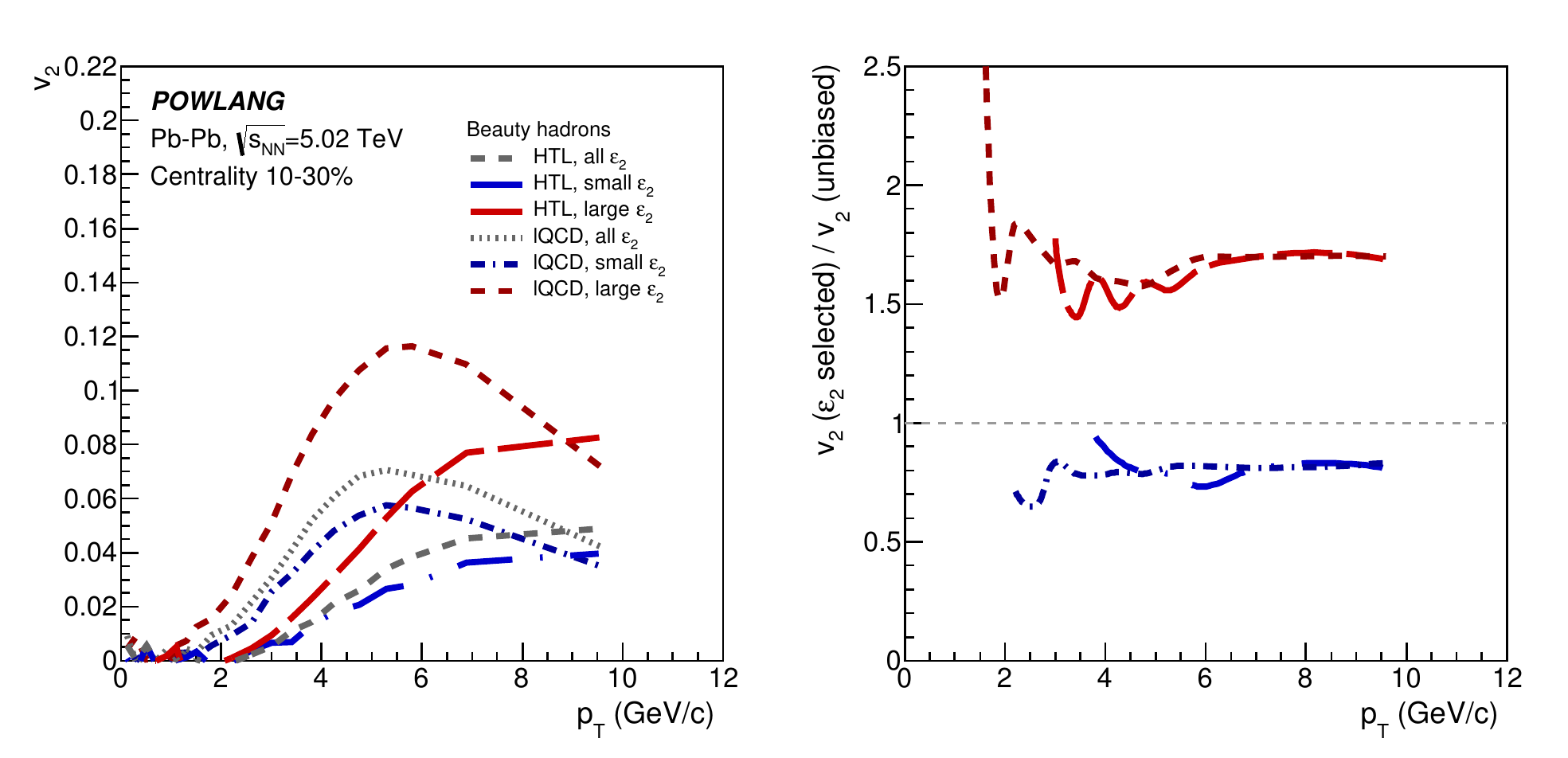}
\includegraphics[clip,width=0.8\textwidth]{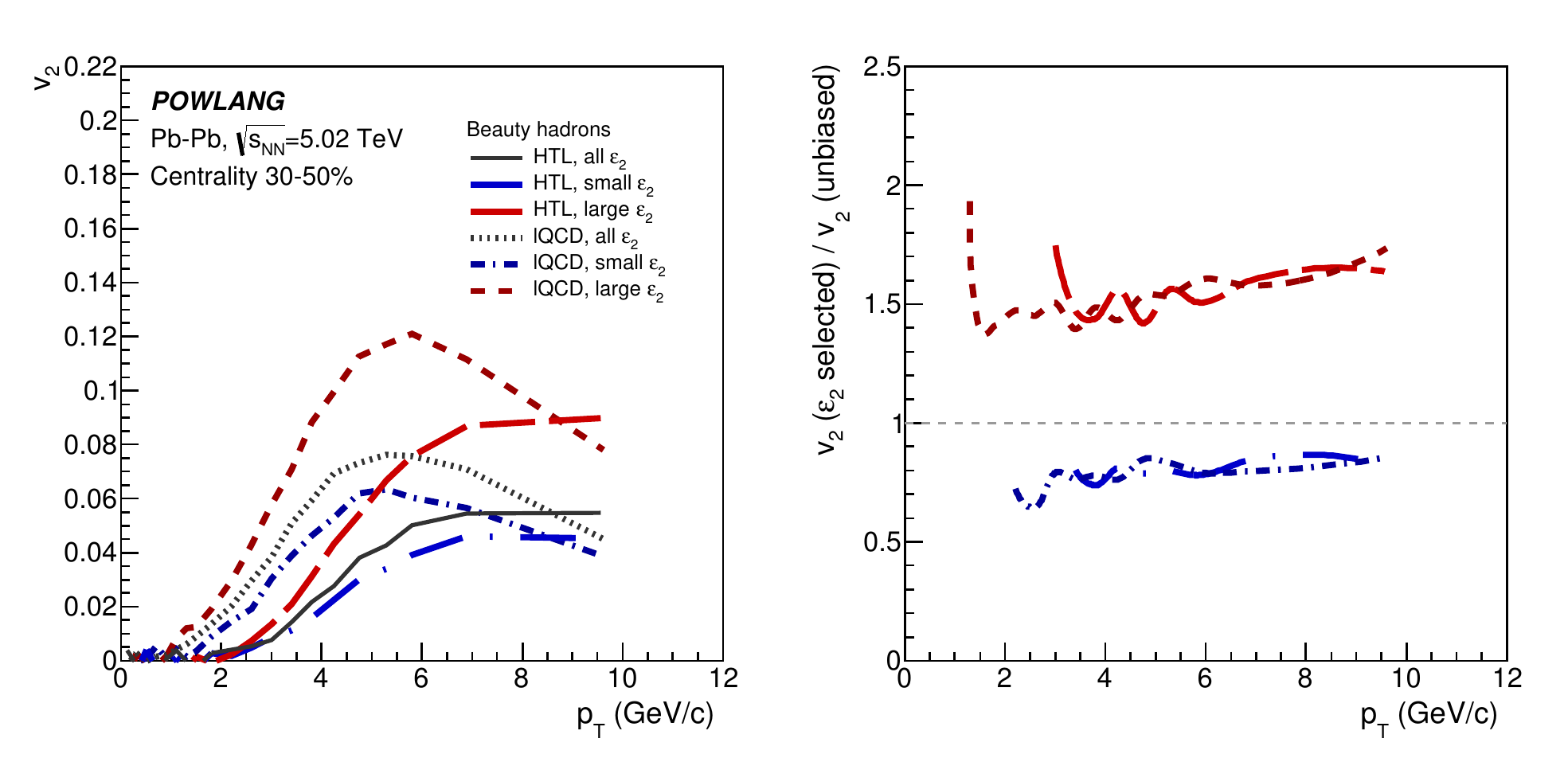}  
\caption{The elliptic-flow coefficient $v_2$ of beauty hadrons in the 0-10\% (top), 10-30\% (middle) and 30-50\% (bottom) most central Pb-Pb collisions at $\sqrt{s_{\rm NN}}\!=\!5.02$ TeV. Results for the 20\% highest-$\epsilon_2$ and to the 60\% lowest-$\epsilon_2$ selection of events are compared to the unbiased case. For both choices of transport coefficients the results display a strong and similar sensitivity to the initial eccentricity.}\label{fig:v2B_sel-vs-mb-all} 
\end{center}
\end{figure}
\begin{figure}[!ht]
\begin{center}
  \includegraphics[clip,width=0.8\textwidth]{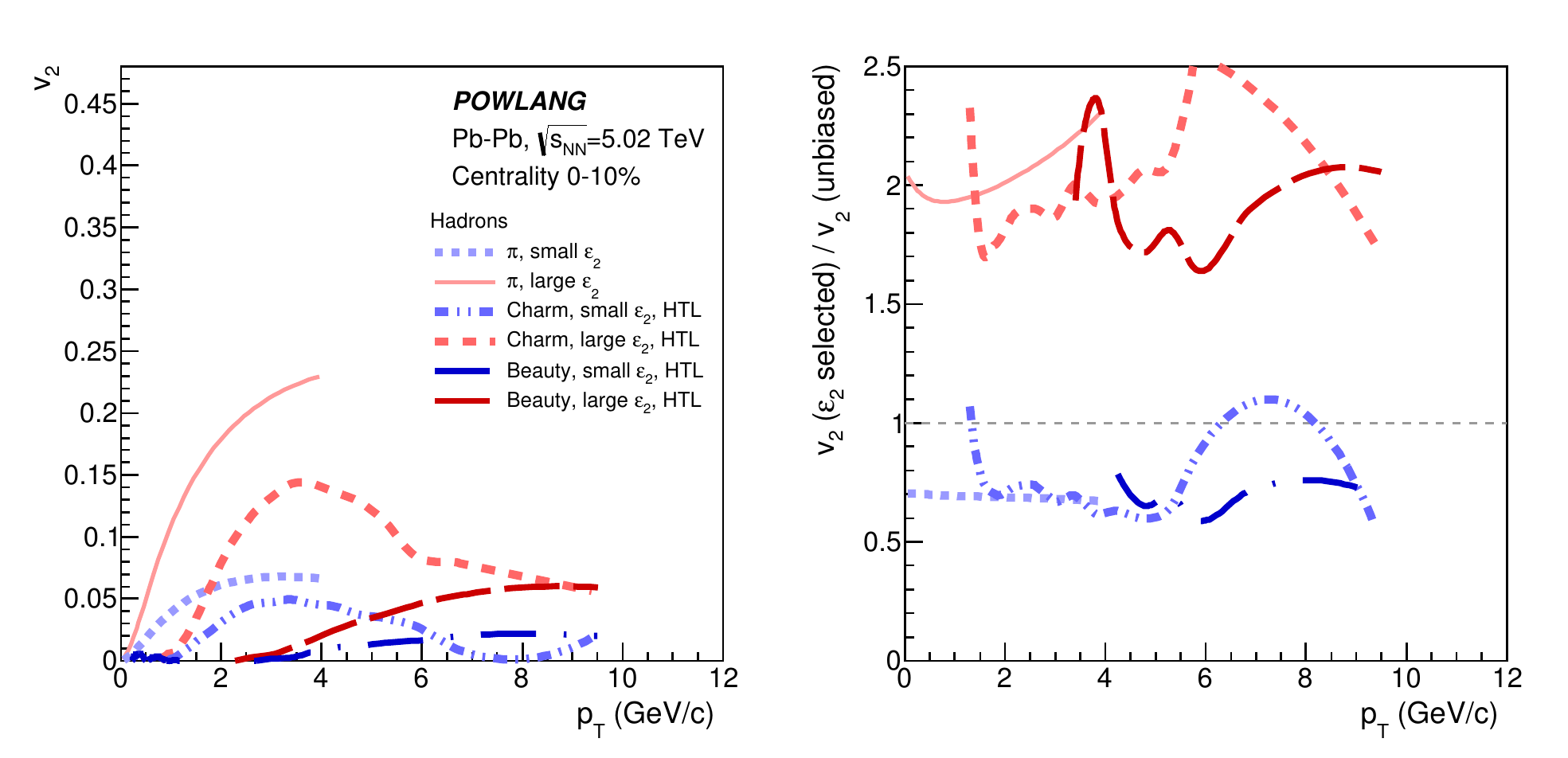}
  \includegraphics[clip,width=0.8\textwidth]{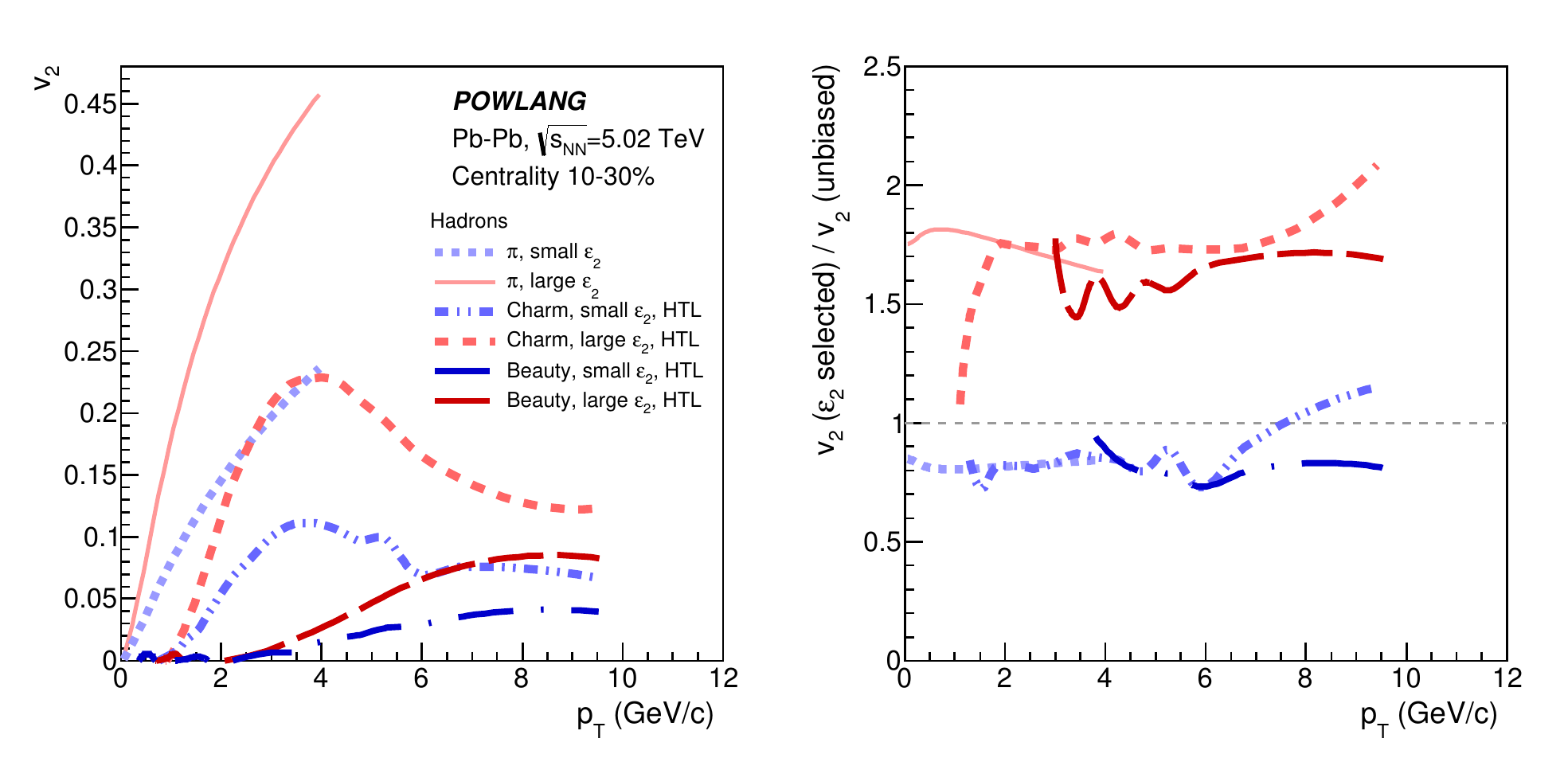}
\includegraphics[clip,width=0.8\textwidth]{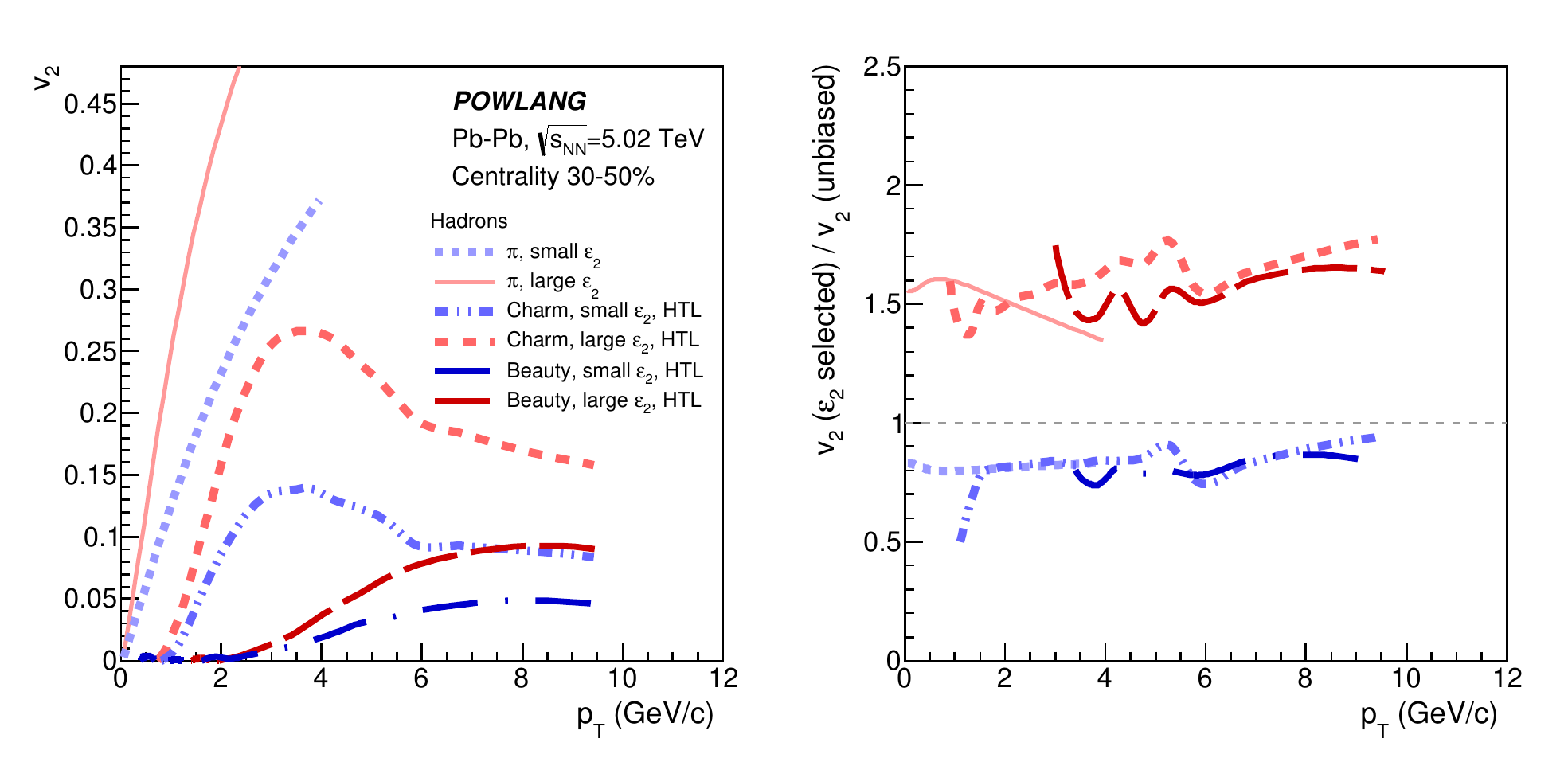}  
\caption{Systematic comparison of the effect of the eccentricity selection on the elliptic flow of light and heavy flavour hadrons in the 0-10\% (top), 10-30\% (middle) and 30-50\% (bottom) most central Pb-Pb collisions at $\sqrt{s_{\rm NN}}\!=\!5.02$ TeV. Deviations from the unbiased case look remarkably similar for pions, charmed and beauty hadrons.}\label{fig:heavy-light} 
\end{center}
\end{figure}
We finally move to consider also beauty quarks and hadrons, focusing on the study of their elliptic flow and comparing the results to the ones found for lighter hadrons. In Fig.~\ref{fig:v2B_sel-vs-mb-all} the $v_2$ coefficients of beauty-hadron distributions obtained selecting the 20\% highest-$\epsilon_2$ and the 60\% lowest-$\epsilon_2$ are shown and compared to the results referring to an unbiased selection of events. The study is performed for the 0-10\%, 10-30\% and 30-50\% centrality classes. Our findings are similar to what already obtained for charm: for all the centrality classes the ratio $v_2^{{\rm high/low}-\epsilon_2}/v_2^{\rm unbiased}$ looks quite constant as a function of the transverse momentum $p_T$ and independent of the choice of the transport coefficients.
In Fig.~\ref{fig:heavy-light} the results for the $v_2$ of beauty hadrons with event-shape-engineering are compared to those for charmed hadrons and pions. The effect of the eccentricity selection is similar for particles with very different masses and the largest deviations from unity of $v_2^{\rm ESE}/v_2^{\rm unbiased}$ are observed in the 0-10\% centrality class. Such a systematic comparison suggests that the quantity $v_2^{{\rm ESE}}/v_2^{\rm unbiased}$ reflects essentially the initial geometric deformation of the system. 

\section{Discussion and perspectives}\label{Sec:conclusions}
Event-shape-engineering studies of particle $p_T$-spectra and flow in relativistic heavy-ion collisions, in which events are organized first in centrality classes and then in subsamples of high/low eccentricity, have the potential to provide a richer information on the produced medium, disentangling the effects of the size and density of the fireball from the ones related to its geometric asymmetry. In this paper we decided to focus on what one can learn in principle applying such a strategy to the study of heavy-flavour observables, showing results obtained with our POWLANG transport setup. In this case, in fact, one deals with external probes -- the charm or beauty quarks -- produced off-equilibrium in hard processes occurring before the formation of a thermalized Quark-Gluon Plasma. These heavy quarks then cross the medium, interacting with its constituents, before hadronizing and being detected. We expect then that the initial density and size of the medium, beside its shape, affect the final momentum and angular distribution of charm and beauty hadrons.  

Notice that, at variance with the actual experimental situation in which an estimator based on the average flow measured in a different kinematic region is used as a a proxy of the initial geometric asymmetry, in our simulations we can really select events on the basis of their initial elliptic o triangular eccentricity. On the other hand experimental analysis can rely on a huge statistics in each centrality class; performing an analogous theoretical study with full event-by-event simulations would require huge computing and storage resources.
Before starting a similar massive campaign it is important to get a solid estimate of the size of the effect one can observe and of what one can learn on the medium and on its interaction with the external probes: this can be done within a simplified approach. For each of the considered subset of collisions we relied then on a one-shot hydrodynamic simulation with a proper average initial condition. Of course, this prevented us from disentangling eccentricity and centrality as cleanly as in the experimental analysis and to study, for instance, correlations among radial, elliptic and triangular flow, but allowed us in any case to get a list of interesting results. 

We started our analysis with the nuclear modification factor of charm hadrons, finding that, within a given centrality class, the selection of events with high/low initial eccentricity does not affect significantly the results. The small effect, at most of order 10-20\%, looks compatible with the positive correlation between eccentricity and impact parameter of the collisions, which entails that more eccentric events are also on average more peripheral, hence leading to a milder quenching of the heavy-quark momentum. As already discussed, experimental analysis try to remove such an artificial correlation performing the selection on eccentricity in very small centrality bins. The small size of the effect (deviations from unity of the ratio of the heavy-flavour $p_T$-distributions in high/low-$\epsilon_2$ events over the unbiased case being small), the current level of precision of the data and the slightly different procedure in performing the eccentricity selection do not allow to draw meaningful conclusions from a  comparison with the present experimental data. However, in the near future, reducing the experimental uncertainties thanks to larger samples of data and performing a cleaner separation of eccentricity and centrality in theory calculations will allow one to extract a reacher information on the heavy-quark interaction with the medium.

On the contrary, a selection based on the event-shape was found to lead to a major effect on the elliptic and triangular flow: we obtained results for the charmed hadron $v_2$ and $v_3$ in high-eccentricity events a factor 2 larger than in the unbiased case. The ratio $v_n^{\rm ESE}/v_n^{\rm unbiased}$ looks quite constant as a function of $p_T$. Interestingly, while results for the $v_2$ and $v_3$ obtained with weak-coupling (HTL curves) or non-perturbative (lQCD curves) transport coefficients display significant differences, the ratio between the high/low-eccentricity results and the unbiased case looks pretty independent of the modeling of the interaction with the medium, suggesting that the effects depends mainly on the initial geometry of the fireball. Also the dependence of $v_n^{\rm ESE}/v_n^{\rm unbiased}$ on centrality is quite weak: for the triangular flow it is completely negligible; in the case of the elliptic flow, deviations from unity of the ratio $v_2^{{\rm high}-\epsilon_2}/v_2^{\rm unbiased}$ tend to slightly decrease moving from central to more peripheral collisions, the smallest effect being observed for charm quarks in the 60-80\% centrality class. This last observation suggests a limited interaction of the heavy quark in the case of a less thick and dense medium, which cannot leave the imprints of its initial geometry in the final angular distribution of charm quarks.

We decided then to follow a complementary strategy, namely to select events of a given initial eccentricity $\epsilon_2$ and $\epsilon_3$ and study how the results for the flow coefficients $v_2$ and $v_3$ change when considering different centrality classes. We started considering light hadrons, coming from the hadronization of the bulk medium. We saw that in the case of soft hadrons decoupling from a freeze-out hypersurface, for a given initial eccentricity, the flow pattern looks essentially the same in the different centrality classes: anisotropies in the particle distributions simply reflect the corresponding asymmetries in the fluid-velocity field at freeze-out, arising from the hydrodynamic response of the medium to its initial geometric deformation. In the case of heavy flavour distributions, however, things are more complicate, since we are not dealing with particles which are part of the bulk medium from the beginning of its evolution, but with hadrons arising from $c$ and $b$ quarks produced in initial hard partonic processes, with momentum distributions described by perturbative-QCD. In this case we expect that the centrality of the collision plays an important role in determining the response of the final particle distributions to the same initial geometric deformation, since a medium of larger size, longer lifetime and higher density should affect more strongly the propagation of the heavy quarks. This is what we actually observed at the quark level, both for the $v_2$ and the $v_3$: selecting events with the same $\epsilon_{2/3}$ we found a larger elliptic/triangular flow of charm quarks in more central collisions. Hadronization, modeled in our scheme via recombination with light thermal partons following the flow of the medium, tends to wash out this difference, although some effect is still visible, in particular in the case of $v_3$. We hope our observations can motivate future experimental analysis along this direction.

Finally we moved to beauty, focusing on its elliptic flow, and our main finding is that, although the $v_2$ of beauty hadrons is quite small, the effect of the eccentricity selection on the azimuthal distributions, once normalized to the unbiased result, turns out to be of the same size of the one of charmed and light hadrons.

Our study presented in this paper must be considered just a first step in the direction of better constraining the heavy-quark interaction with the medium and the response to the event-by-event fluctuations in the initial state of the latter. In the future we can certainly improve our results, employing and event-by-event approach allowing a study of all possible correlations of the various kind of flow (radial, elliptic and triangular) among themselves and with the fluctuations of the initial geometry of the medium. This, however, will be a very demanding task from the point of view of computing time and storage resources. We believe that this first cheaper exploratory study has already been able to provide some interesting indications motivating future, more refined, ESE-analysis addressing for instance the triangular flow and more peripheral centrality classes. It also permits a first comparison with the experimental data, which in the following years -- with increasing statistics -- will become more precise.

\bibliography{paper}

\providecommand{\newblock}{}
\begin{thebibliography}{10}
\expandafter\ifx\csname url\endcsname\relax
  \def\url#1{{\tt #1}}\fi
\expandafter\ifx\csname urlprefix\endcsname\relax\def\urlprefix{URL }\fi
\providecommand{\eprint}[2][]{\url{#2}}

\bibitem{Adare:2010de}
Adare A {\em et~al.\/} (PHENIX) 2011 {\em Phys. Rev.\/} {\bf C84} 044905
  (\textit{Preprint} \eprint{1005.1627})

\bibitem{Abelev:2006db}
Abelev B~I {\em et~al.\/} (STAR) 2007 {\em Phys. Rev. Lett.\/} {\bf 98} 192301
  [Erratum: Phys. Rev. Lett.106,159902(2011)] (\textit{Preprint}
  \eprint{nucl-ex/0607012})

\bibitem{ALICE:2012ab}
Abelev B {\em et~al.\/} (ALICE) 2012 {\em JHEP\/} {\bf 09} 112
  (\textit{Preprint} \eprint{1203.2160})

\bibitem{Adamczyk:2014uip}
Adamczyk L {\em et~al.\/} (STAR) 2014 {\em Phys. Rev. Lett.\/} {\bf 113} 142301
  (\textit{Preprint} \eprint{1404.6185})

\bibitem{Sirunyan:2017xss}
Sirunyan A~M {\em et~al.\/} (CMS) 2018 {\em Phys. Lett.\/} {\bf B782} 474--496
  (\textit{Preprint} \eprint{1708.04962})

\bibitem{Abelev:2013lca}
Abelev B {\em et~al.\/} (ALICE) 2013 {\em Phys. Rev. Lett.\/} {\bf 111} 102301
  (\textit{Preprint} \eprint{1305.2707})

\bibitem{Adamczyk:2017xur}
Adamczyk L {\em et~al.\/} (STAR) 2017 {\em Phys. Rev. Lett.\/} {\bf 118} 212301
  (\textit{Preprint} \eprint{1701.06060})

\bibitem{Sirunyan:2017plt}
Sirunyan A~M {\em et~al.\/} (CMS) 2018 {\em Phys. Rev. Lett.\/} {\bf 120}
  202301 (\textit{Preprint} \eprint{1708.03497})

\bibitem{Francis:2015daa}
Francis A, Kaczmarek O, Laine M, Neuhaus T and Ohno H 2015 {\em Phys. Rev.\/}
  {\bf D92} 116003 (\textit{Preprint} \eprint{1508.04543})

\bibitem{CaronHuot:2008uh}
Caron-Huot S and Moore G~D 2008 {\em JHEP\/} {\bf 02} 081 (\textit{Preprint}
  \eprint{0801.2173})

\bibitem{Xu:2017hgt}
Xu Y, Nahrgang M, Bernhard J~E, Cao S and Bass S~A 2017 {\em Nucl. Phys.\/}
  {\bf A967} 668--671 (\textit{Preprint} \eprint{1704.07800})

\bibitem{Rapp:2018qla}
Beraudo A {\em et~al.\/} 2018 {\em Nucl. Phys.\/} {\bf A979} 21--86
  (\textit{Preprint} \eprint{1803.03824})

\bibitem{Policastro:2001yc}
Policastro G, Son D~T and Starinets A~O 2001 {\em Phys. Rev. Lett.\/} {\bf 87}
  081601 (\textit{Preprint} \eprint{hep-th/0104066})

\bibitem{Adam:2015jda}
Adam J {\em et~al.\/} (ALICE) 2016 {\em JHEP\/} {\bf 03} 082 (\textit{Preprint}
  \eprint{1509.07287})

\bibitem{Zhou:2017ikn}
Zhou L (STAR) 2017 {\em Nucl. Phys.\/} {\bf A967} 620--623 (\textit{Preprint}
  \eprint{1704.04364})

\bibitem{Sirunyan:2017oug}
Sirunyan A~M {\em et~al.\/} (CMS) 2017 {\em Phys. Rev. Lett.\/} {\bf 119}
  152301 (\textit{Preprint} \eprint{1705.04727})

\bibitem{Adare:2012yxa}
Adare A {\em et~al.\/} (PHENIX) 2012 {\em Phys. Rev. Lett.\/} {\bf 109} 242301
  (\textit{Preprint} \eprint{1208.1293})

\bibitem{Abelev:2014hha}
Abelev B~B {\em et~al.\/} (ALICE) 2014 {\em Phys. Rev. Lett.\/} {\bf 113}
  232301 (\textit{Preprint} \eprint{1405.3452})

\bibitem{Adam:2016ich}
Adam J {\em et~al.\/} (ALICE) 2016 {\em Phys. Rev.\/} {\bf C94} 054908
  (\textit{Preprint} \eprint{1605.07569})

\bibitem{Sirunyan:2018toe}
Sirunyan A~M {\em et~al.\/} (CMS) 2018 {\em Phys. Rev. Lett.\/} {\bf 121}
  082301 (\textit{Preprint} \eprint{1804.09767})

\bibitem{Beraudo:2015wsd}
Beraudo A, De~Pace A, Monteno M, Nardi M and Prino F 2016 {\em JHEP\/} {\bf 03}
  123 (\textit{Preprint} \eprint{1512.05186})

\bibitem{Singha:2018cdj}
Singha S (STAR) 2018 {\em {27th International Conference on Ultrarelativistic
  Nucleus-Nucleus Collisions (Quark Matter 2018) Venice, Italy, May 14-19,
  2018}\/} (\textit{Preprint} \eprint{1807.04771})

\bibitem{Chatterjee:2017ahy}
Chatterjee S and Bożek P 2018 {\em Phys. Rev. Lett.\/} {\bf 120} 192301
  (\textit{Preprint} \eprint{1712.01189})

\bibitem{Das:2016cwd}
Das S~K, Plumari S, Chatterjee S, Alam J, Scardina F and Greco V 2017 {\em
  Phys. Lett.\/} {\bf B768} 260--264 (\textit{Preprint} \eprint{1608.02231})

\bibitem{Nahrgang:2014vza}
Nahrgang M, Aichelin J, Bass S, Gossiaux P~B and Werner K 2015 {\em Phys.
  Rev.\/} {\bf C91} 014904 (\textit{Preprint} \eprint{1410.5396})

\bibitem{Beraudo:2017gxw}
Beraudo A, De~Pace A, Monteno M, Nardi M and Prino F 2018 {\em JHEP\/} {\bf 02}
  043 (\textit{Preprint} \eprint{1712.00588})

\bibitem{Schukraft:2012ah}
Schukraft J, Timmins A and Voloshin S~A 2013 {\em Phys. Lett.\/} {\bf B719}
  394--398 (\textit{Preprint} \eprint{1208.4563})

\bibitem{Adam:2015eta}
Adam J {\em et~al.\/} (ALICE) 2016 {\em Phys. Rev.\/} {\bf C93} 034916
  (\textit{Preprint} \eprint{1507.06194})

\bibitem{Beraudo:2018bxb}
Beraudo A, De~Pace A, Monteno M, Nardi M and Prino F 2018 {\em {27th
  International Conference on Ultrarelativistic Nucleus-Nucleus Collisions
  (Quark Matter 2018) Venice, Italy, May 14-19, 2018}\/} (\textit{Preprint}
  \eprint{1807.03180})

\bibitem{Acharya:2018bxo}
Acharya S {\em et~al.\/} (ALICE) 2018  (\textit{Preprint} \eprint{1809.09371})

\bibitem{Prado:2016szr}
Prado C~A~G, Noronha-Hostler J, Katz R, Suaide A~A~P, Noronha J, Munhoz M~G and
  Cosentino M~R 2017 {\em Phys. Rev.\/} {\bf C96} 064903 (\textit{Preprint}
  \eprint{1611.02965})

\bibitem{Christiansen:2016uaq}
Christiansen P 2016 {\em J. Phys. Conf. Ser.\/} {\bf 736} 012023
  (\textit{Preprint} \eprint{1606.07963})

\bibitem{Gossiaux:2017nwz}
Gossiaux P~B, Aichelin J, Nahrgang M, Ozvenchuk V and Werner K 2017 {\em Nucl.
  Phys.\/} {\bf A967} 672--675 (\textit{Preprint} \eprint{1705.02271})

\bibitem{DelZanna:2013eua}
Del~Zanna L, Chandra V, Inghirami G, Rolando V, Beraudo A, De~Pace A, Pagliara
  G, Drago A and Becattini F 2013 {\em Eur. Phys. J.\/} {\bf C73} 2524
  (\textit{Preprint} \eprint{1305.7052})

\bibitem{Qiu:2011iv}
Qiu Z and Heinz U~W 2011 {\em Phys. Rev.\/} {\bf C84} 024911 (\textit{Preprint}
  \eprint{1104.0650})

\bibitem{Alberico:2013bza}
Alberico W~M, Beraudo A, De~Pace A, Molinari A, Monteno M, Nardi M, Prino F and
  Sitta M 2013 {\em Eur. Phys. J.\/} {\bf C73} 2481 (\textit{Preprint}
  \eprint{1305.7421})

\bibitem{Banerjee:2011ra}
Banerjee D, Datta S, Gavai R and Majumdar P 2012 {\em Phys. Rev.\/} {\bf D85}
  014510 (\textit{Preprint} \eprint{1109.5738})

\bibitem{Beraudo:2014boa}
Beraudo A, De~Pace A, Monteno M, Nardi M and Prino F 2015 {\em Eur. Phys. J.\/}
  {\bf C75} 121 (\textit{Preprint} \eprint{1410.6082})

\bibitem{Cao:2018ews}
Cao S {\em et~al.\/} 2018  (\textit{Preprint} \eprint{1809.07894})

\end{thebibliography}

\end{document}